\documentclass[%
 reprint,
nofootinbib,
 amsmath,amssymb,
 aps,
aps,prd,10pt,twocolumn,amsmath,superscriptaddress,amssymb,showpacs,floatfix,nofootinbib,longbibliography]{revtex4-2}
\usepackage{graphicx}
\usepackage{dcolumn}
\usepackage{bm}
\usepackage{comment}
\usepackage{hyperref}
\usepackage{amsfonts,amsmath,amssymb,bm,bbm,mathrsfs}
\usepackage{color}
\usepackage{url}
\usepackage{float}
\usepackage{slashed}
\usepackage{enumitem}
\usepackage{leftindex}
\usepackage[capitalise]{cleveref}
\usepackage{braket}
\usepackage{bbold}
\usepackage{multirow}

\hypersetup{
    pdfnewwindow=true,      
    colorlinks=true,       
    linkcolor=blue,          
    citecolor=blue,        
    filecolor=blue,      
    urlcolor=blue        
}

\newcommand{\orcid}[1]{\begingroup
  \!\hypersetup{hidelinks}\href{https://orcid.org/#1}{\includegraphics[width=10pt]{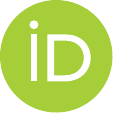}\!\!} \endgroup}

\usepackage{hyperref}
\usepackage{xcolor}

\newcommand{\s}{{\rm s}}
\newcommand{\dd}{{\rm d}}
\newcommand{\X}{{\rm X}}

\begin{document}

\preprint{ULB-TH-XXXX}

\title{21\,cm Cosmology Sensitivity to the Primordial Power Spectrum }

\author{Justus Schwagereit \orcid{0009-0001-9987-1433}\,}
\email{justus.roman.schwagereit@ulb.be}
\affiliation{Service de Physique Th\'eorique, \href{https://ror.org/01r9htc13}{ Universit\'e Libre de Bruxelles}, B-1050 Brussels, Belgium}

\author{Laura Lopez-Honorez \orcid{0000-0002-4149-3935}\,}
\email{laura.lopez.honorez@ulb.be}
\affiliation{Service de Physique Th\'eorique, \href{https://ror.org/01r9htc13}{ Universit\'e Libre de Bruxelles}, B-1050 Brussels, Belgium}
\affiliation{Theoretische Natuurkunde \& The International Solvay Institutes, \href{https://ror.org/006e5kg04}{Vrije Universiteit Brussel}, B-1050 Brussels, Belgium}

\author{Vivian Poulin \orcid{0000-0002-9117-5257}\,}
\email{vivian.poulin@umontpellier.fr}
\affiliation{\href{https://ror.org/00nrbzg90}{Laboratoire Univers et Particules de Montpellier} (LUPM), CC 72, Place Eug\`ene Bataillon, 34095 Montpellier -- Cedex 5, France}

\date{\today}

\begin{abstract}
Deviations from a power-law primordial power spectrum (PPS), such as a running of the scalar spectral index and a running thereof, arise in a variety of inflationary scenarios. We forecast the constraining power of future experiments that probe the 21\,cm power spectrum arising from Cosmic Dawn and the Epoch of Reionisation to such models. We find that combining HERA with Planck data improves projected constraints on $\alpha_\s$ and $\beta_\s$ by about 25\% relative to Planck alone. Once Simons Observatory (SO) data are included, they dominate the constraint on $\alpha_\s$, while HERA still improves the constraint on $\beta_\s$. We further show that, when allowing the effective number of relativistic species, $N_{\rm eff}$, to vary, HERA can significantly break degeneracies between the PPS parameters and $N_{\rm eff}$ that exist in CMB data. We compare our HERA forecasts with those that we obtain for the final {\tt AA4} configuration of SKA and find improvements on the uncertainties on $\alpha_\s$, $\beta_\s$ and of the optical depth to reionization with SKA. Finally, we assess the impact of modelling uncertainties and the halo mass function parametrization, finding our conclusions to be qualitatively robust. 
\end{abstract}



\maketitle

\section{\label{sec:level1}Introduction}

The standard model of cosmology, $\Lambda$CDM, contains six free parameters, two of which are used to characterise the primordial power spectrum of curvature perturbations (PPS) generated by inflation: the spectrum amplitude $A_\s$ and tilt $n_\s$. However, departures from this simple power-law are conceivable: slow-roll single-field inflation generally predicts a small running of the tilt with wavenumber $k$, $|\alpha_\s|\sim 10^{-3}$  with $\alpha_\s:=\dd n_\s/\dd\ln k$~\cite{Kosowsky:1995aa, CMBPolStudyTeam:2008rgp}. 

In the absence of higher-order runnings, the Planck collaboration set a constraint of $\alpha_\s=-0.0045\pm0.0067$~\cite{Planck:2018jri} from observations of the cosmic microwave background (CMB). The ACT collaboration tightened this to $\alpha_\s=0.0062\pm0.0052$~\cite{AtacamaCosmologyTelescope:2025nti}. In a joint fit of both $\alpha_\s$ and the running of $\alpha_\s$, $\beta_\s:=\dd\alpha_\s/\dd\ln k$, Planck set a constraint of $\alpha_\s=0.002\pm0.010$, $\beta_\s=0.010\pm0.013$. Reference~\cite{Fairbairn:2025fko} used Planck+ACT data to obtain $\alpha_\s=0.0120\pm0.0062$, $\beta_\s=0.0186\pm0.0096$, Planck+ACT+SPT data to obtain $\alpha_\s=0.0080\pm0.0057$, $\beta_\s=0.0148\pm0.0088$ and Planck+ACT+SPT+eBOSS Lyman-$\alpha$ forest data to obtain $\alpha_\s=-0.0032\pm0.0039$, $\beta_\s=-0.0076\pm0.0035$, although the latter values may be subject to systematics related to Lyman-$\alpha$ measurements.

It has been pointed out recently that CMB bounds on $N_{\rm eff}$, the effective number of fermionic relativistic degrees of freedom, strongly depend on assumptions on the PPS~\cite{Garny:2026gcs}. Concretely, it was shown that values of $\Delta N_{\rm eff}=N_{\rm eff}-3.044\gtrsim0.5$ could be reconciled with CMB data by allowing for non-zero runnings $\alpha_\s\sim\beta_\s\sim0.02$. Here, $3.044$ is the standard model prediction for $N_{\rm eff}$~\cite{Froustey:2020mcq, Bennett:2020zkv}. Such values of $\Delta N_{\rm eff}$ could alleviate the Hubble tension~\cite{Schoneberg:2021qvd} but, in the absence of modifications to the primordial power spectrum, are disfavoured by large-$\ell$ CMB observations as they lead to a damping of the large-$\ell$ tail of the CMB TT power spectrum. This damping can be counteracted by allowing for non-zero positive values of $\alpha_\s$ and $\beta_\s$ (see Fig.~\ref{fig:cmb_tcl}). In turn, constraints on $\alpha_\s$ and $\beta_\s$ from CMB observations get significantly relaxed when allowing for non-zero values of $\Delta N_{\rm eff}$. It is therefore useful to derive additional constraints on both the PPS and $\Delta N_{\rm eff}$ from lower-redshift large-scale structure observations, like Lyman-$\alpha$ data~\cite{Fairbairn:2025fko}, the Euclid 3x2pt probe~\cite{Euclid:2025hlc}, or upcoming 21\,cm observations, and test whether such probes are able to break this degeneracy.

A non-zero value of the running $\alpha_\s\sim-0.01$ has previously been invoked to solve tension between the amplitude and tilt of the matter power spectrum as inferred from Planck data and eBOSS Lyman-$\alpha$ forest measurements~\cite{Rogers:2023upm}.  Specifically, eBOSS data preferred a lower amplitude and stronger tilt of the matter power spectrum at $k=0.009\,{\rm s\,km}^{-1}\approx1\,h\,{\rm Mpc}^{-1}$ and $z=3$ than one obtains when extrapolating the tilt and amplitude of the PPS from Planck measurements at smaller $k$. The analysis further assumed that  the transfer function, $T(k,z)$ -~t hat connects the PPS to the matter power spectrum ~- is fixed, and the PPS follows a power law. The resulting tension between the amplitude and tilt of the matter power spectrum amounted to $5\,\sigma$ and could be resolved by a model in which $\alpha_\s=-0.01$, which remains compatible with Planck data.
Note  however that, more recently, Ref.~\cite{Fairbairn:2025fko} re-analysed the same data using the more recent likelihood of~\cite{Fernandez:2023grg}, which adopts a more conservative modelling of astrophysical effects and uncertainties, and saw the preference for non-zero $\alpha_\s$ (or non-zero $\alpha_\s$ and $\beta_\s$) disappear. The reason for this was that the likelihood of~\cite{Fernandez:2023grg} prefers a lower amplitude of the matter power spectrum than would be expected from Planck measurements and $\Lambda$CDM cosmology, but not a stronger tilt, a feature that cannot be accommodated by the $\Lambda$CDM+$\alpha_\s$ or $\Lambda$CDM+$\alpha_\s$+$\beta_\s$ models. 

Here we assess how well upcoming cosmological probes, alone and in combination with each other, can detect non-zero values of runnings of $n_\s$. Specifically, we consider synergies between \textit{Planck} data~\cite{Planck:2018vyg,Planck:2018jri}, the future CMB experiment Simons Observatory (SO)~\cite{SimonsObservatory:2018koc, SimonsObservatory:2025wwn} and $21\,$cm cosmology experiments that probe inhomogeneities in the intergalactic medium, namely the Hydrogen Epoch of Reionization Array (HERA)~\cite{HERA:2021bsv,HERA:2025ajm} and the Square Kilometre Array (SKA)~\cite{Dewdney:2009tmd,Mellema:2012ht}. In the latter experiments, the exact form of the PPS is expected to affect the  21\,cm signal through its impact on the distribution of haloes that host the galaxies shaping the signal. In our approach, this effect is captured  by the halo mass function (HMF), which is a measure of the abundance of haloes as a function of their mass.

Previous works have investigated the ability of high-redshift 21\,cm power spectrum observations to probe the small-scale matter power spectrum, see e.g.~\cite{Sitwell:2013fpa,Munoz:2019hjh,Munoz:2020mue,Jones:2021mrs,Flitter:2022pzf,Giri:2022nxq,Minoda:2022nso,Dandoy:2025twl,Facchinetti:2026wed}, as well as running of the scale dependence of the primordial power spectrum~\cite{Mao:2008ug,Adshead:2010mc,Shimabukuro:2014ava,Kohri:2013mxa,Sekiguchi:2017cdy,Minoda:2022nso}. We improve upon those works using more advanced modelling of the 21\,cm signal with self-consistent, semi-analytical simulations, propagation through the HMF and marginalisation over astrophysical parameters, as well as by considering the interplay between CMB and 21\,cm data. Furthermore some analysis have considered constraints on the scale dependence of the PPS from the reionisation history computed using {\tt 21cmFAST}~\cite{Minoda:2023oqs}, UV luminosity functions~\cite{Yoshiura:2020soa} and the 21\,cm global signal~\cite{Yoshiura:2018zts,Yoshiura:2019zxq,Yoshiura:2023xkd}.

This paper is organized as follows. After a brief introduction to the formalism used in this paper in Sec.~\ref{sec:formalism}, we detail  the datasets and the methodology followed in the Fisher matrix forecast analysis in Secs.~\ref{sec:methodology_datasets} and ~\ref{sec:methodology}. We present our results in Sec.~\ref{sec:results} and conclude in Sec.~\ref{sec:conclusion}.
In what follows, we assume Planck 2018 best-fit parameters~\cite{Planck:2018vyg}, which are given in Tab.~\ref{tab:cosmo_parameters}. Unless stated otherwise, confidence intervals are given at $68\%$ confidence level (C.L.). 

\begin{table}[t]
    \centering
    \begin{tabular}{|c|c|c|}
        \hline
        \textbf{Parameter} & \textbf{Fiducial Value} & \textbf{Variation (21\,cm)} \\
        \hline
        $\omega_{\rm b}$ & 0.02242 & $\pm0.4\%$ \\
        $\omega_{\rm cdm}$ & 0.11933 & $\pm0.15\%$ \\
        $h$ & 0.6766 & $\pm5\%$ \\
        $\tau$ & 0.0561 & -- \\
        $\ln(10^{10}\,A_\s)$ & 3.047 & $\pm0.1\%$ \\
        $n_\s$ & 0.9665 & $\pm0.1\%$ \\
        $\alpha_\s$ & 0.0 & $\pm0.0005$ \\
        $\beta_\s$ & 0.0 & $\pm0.0004$ \\
        $\Delta N_{\rm eff}$ & 0.0 & $\pm0.015$ \\
        \hline
    \end{tabular}
    \caption{Fiducial values of the cosmological parameters considered, as well as their respective variations in the HERA $21\,$cm cosmology Fisher forecast. When shown, the $\%$ sign refers to the fiducial value. $\tau$ is only an input parameter in the CMB forecasts, not the $21\,$cm forecast. Here, $\Delta N_{\rm eff}:=N_{\rm eff}-3.044$.}
    \label{tab:cosmo_parameters}
\end{table}

\section{\label{sec:formalism}Formalism}
In this section, we introduce how we characterise the primordial power spectrum as well as the astrophysics that shape the $21\,$cm signal.

\subsection{Primordial Power Spectrum}
\label{sec:PS}

\begin{figure*}
    \centering
    \includegraphics[width=0.49\linewidth]{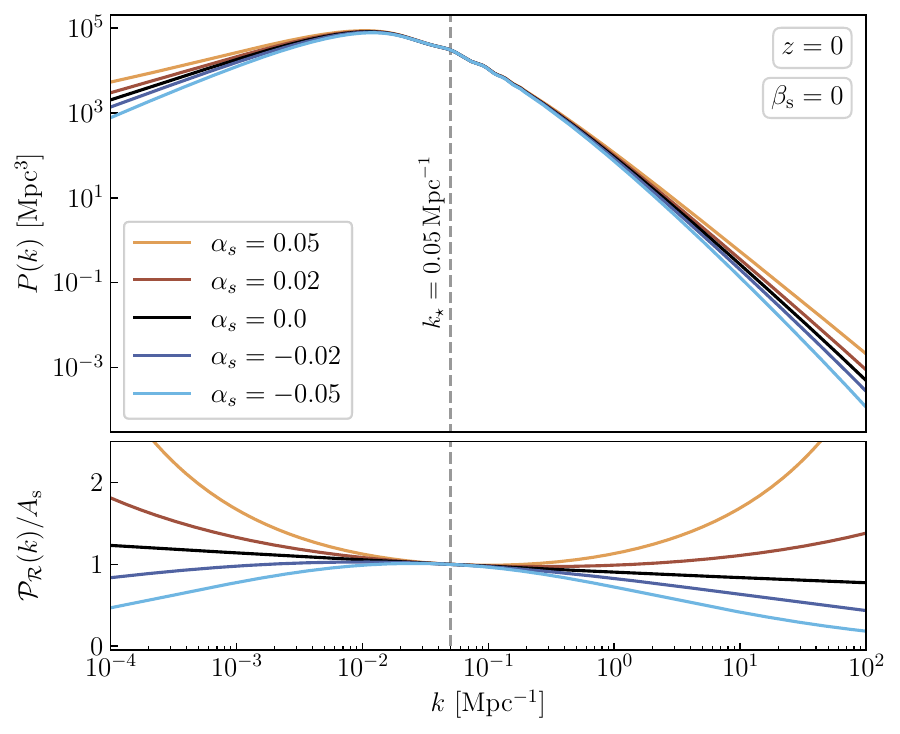}
    \includegraphics[width=0.49\linewidth]{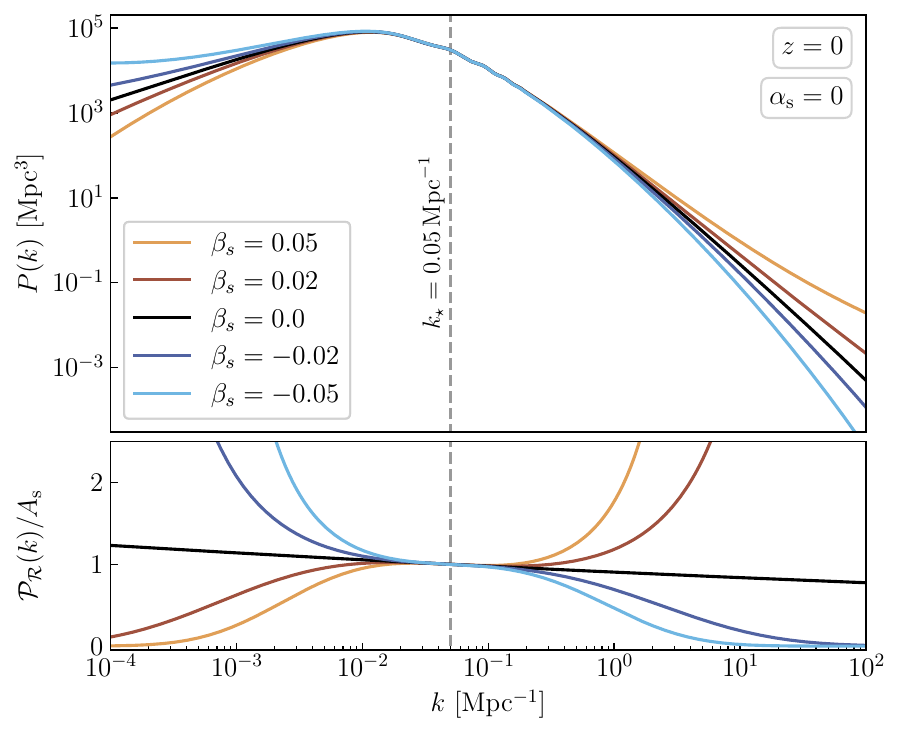}
    \caption{{\bf Impact of the running on the linear matter power spectrum.} \textit{Left}: Impact of variations of $\alpha_\s$ on the linear matter power spectrum at redshift $0$ (\textit{upper panel}) and the primordial power spectrum of curvature perturbations (\textit{lower panel}), for a fixed $\beta_\s=0$. \textit{Right}: Same for variations of $\beta_\s$ and a fixed $\alpha_\s=0$. The dashed vertical line indicates the position of the pivot scale $k_\star$. }
    \label{fig:pk_alphas_betas}
\end{figure*}

The standard model of cosmology assumes a parametrisation of the power spectrum of primordial scalar curvature perturbations as
\begin{align}
    \mathcal{P}_0(k)=A_\s\left(\frac{k}{k_\star}\right)^{n_\s-1}\,,
    \label{eq:lcdm_primordial_power_spectrum}
\end{align}
where $A_\s$ and $n_\s$ are free parameters and the pivot scale $k_\star$ is usually fixed to $0.05\,{\rm Mpc}^{-1}$, approximately in the logarithmic centre of the range of scales probed by {\it Planck}~\cite{Planck:2018jri}. It is, however, conceivable that the tilt may itself depend on $k$, in which case one may perform a Taylor expansion of $\ln\mathcal P_{\mathcal R}$ in $\ln k$ around $\ln k_\star$. This leads to a PPS, ${\cal P}_{\cal R} (k)$, of the form
\begin{align}
    \ln\mathcal{P}_\mathcal{R}(k)=&\ln\mathcal{P}_0(k)+\frac{1}{2}\alpha_\s\ln^2\left(\frac{k}{k_\star}\right) \nonumber\\
    &+\frac{1}{6}\beta_\s\ln^3\left(\frac{k}{k_\star}\right)+...\,.
    \label{eq:as_bs_primordial_power_spectrum}
\end{align}

Figure~\ref{fig:pk_alphas_betas} shows the impact of the parameters $\alpha_\s$ and $\beta_\s$ on the linear matter power spectrum at redshift zero (upper panels) as well as on the primordial power spectrum  of curvature perturbations (lower panels). 
Because the linear matter power spectrum is obtained from the product of the PPS with the transfer function squared, $P(k,z)\propto T^2(k,z)\times\mathcal{P}_\mathcal{R}(k)$, any modification of the PPS propagates directly to the linear matter power spectrum. At the redshifts of interest, the redshift dependent transfer function can be written as $T(k,z)=T(k)D(z)$, where $D(z)$ is the scale-independent, linear growth factor that captures the redshift-dependence of the linear matter perturbations, $D(z)=\delta_m(z)/\delta_m(z=0)$.
In the left (right) panel, $\beta_s$ ($\alpha_s$) is set to zero while the colored lines show the impact of variations of $\alpha_s$  ($\beta_s$) at the $10^{-2}$ level.  Positive (negative) values of $\alpha_\s$ lead to an increase (suppression) of the amplitude of the PPS for $k$ above and below the pivot scale w.r.t. the case of no running. In contrast, a positive (negative) $\beta_\s$ increases power at large (small) $k$ while suppressing power at $k$ smaller (larger) than the pivot scale. This is simply a consequence of the even or odd power of $\ln(k/k_\star)$ in the expansion in Eq.~(\ref{eq:as_bs_primordial_power_spectrum}).

\begin{figure*}[t]
    \centering
    \includegraphics[width=0.49\linewidth]{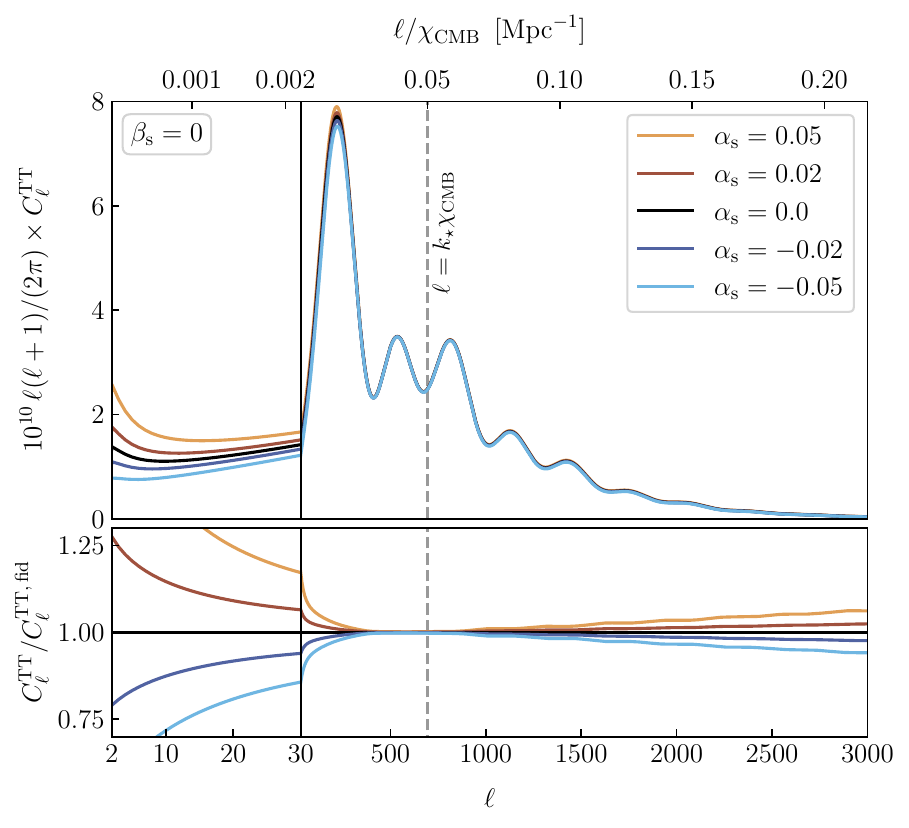}
    \includegraphics[width=0.49\linewidth]{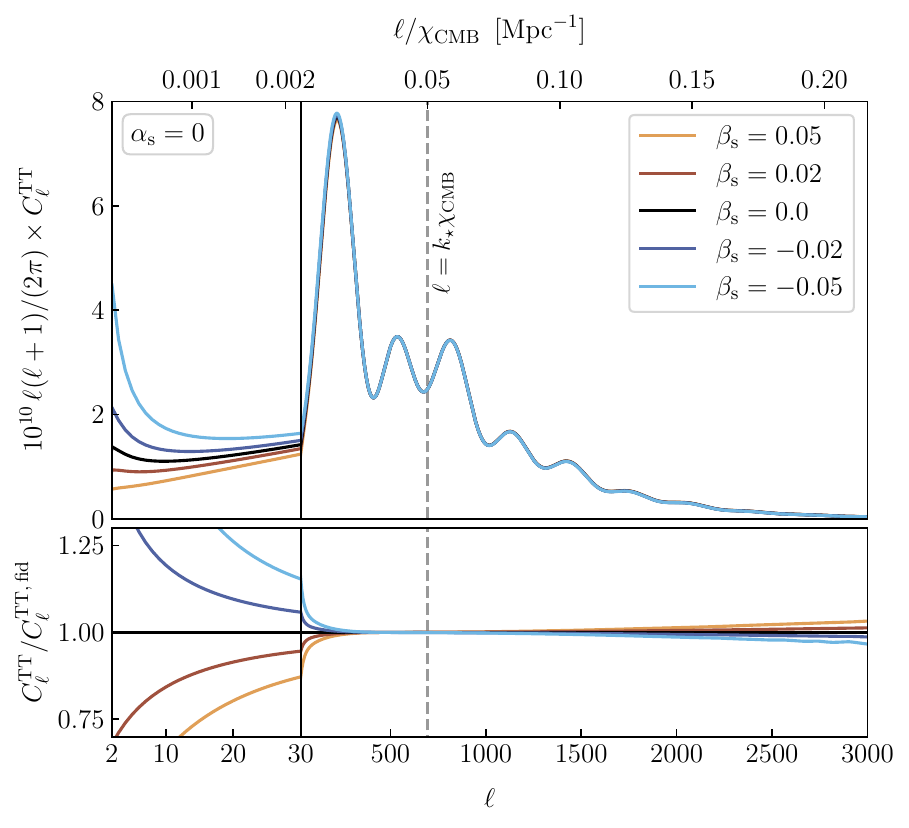}
    \caption{{\bf Impact of the running on the temperature anisotropy power spectrum.} \textit{Left}: Impact of variations of $\alpha_\s$ on the $C_\ell^{\rm TT}$ of the CMB for fixed $\beta_\s=0$. \textit{fid} denotes the fiducial model, $\alpha_\s=\beta_\s=0$. The $C_\ell^{\rm TT}$ are shown in the upper plot, while the lower plot shows the ratio $C_\ell^{\rm TT}(\alpha_\s)/C_\ell^{\rm TT}(\alpha_\s=0)$. Note that $2\leq\ell\leq30$ and $30\leq\ell\leq3000$ use different scalings, the transition between the two regimes is shown as a black vertical line. The approximate correspondence between each $\ell$ and the wavenumber of the PPS that it is most sensitive to, $\ell\sim k\times\chi_{\rm CMB}$, is also shown, where $\chi_{\rm CMB}$ is the comoving distance to the CMB. The dashed vertical line indicates the position of the multipole corresponding to the pivot scale $k_\star$. \textit{Right}: Same for variations of $\beta_\s$ and a fixed $\alpha_\s=0$.}
    \label{fig:cmb_tcl}
\end{figure*}

Similarly, Fig.~\ref{fig:cmb_tcl} shows the impact of the parameters $\alpha_\s$ and $\beta_\s$ on the temperature multipoles of the CMB, $C_\ell^{\rm TT}$. The same behaviour as for the curvature and matter power spectra can be observed, with $\alpha_\s$ and $\beta_\s$ affecting the low-multipole spectrum in opposite directions.
Also shown is the approximate correspondence $k\sim\ell/\chi_{\rm CMB}$, where $\chi_{\rm CMB}\simeq13.9\,{\rm Gpc}$ is the comoving distance to the CMB. Note that the Planck experiment probes the inflationary power spectrum best on scales $10^{-3}\lesssim k/{\rm Mpc}^{-1}\lesssim10^{-1}$~\cite{Planck:2018jri}.

\subsection{Halo Mass Function}
\label{sec:HMF}

To perform simulations of the $21\,$cm signal, we use the semi-numerical code {\tt exo21cmFAST}\footnote{\url{https://github.com/gaetanfacchinetti/exo21cmFAST}}~\cite{Facchinetti:2023slb}, which is forked from {\tt 21cmFAST}\footnote{\url{https://github.com/21cmfast/21cmFAST}}~\cite{Mesinger:2010ne,Murray:2020trn} {\tt v3} and interfaced with the  Boltzmann code {\tt CLASS}\footnote{\url{https://github.com/lesgourg/class_public/}}~\cite{Lesgourgues:2011re,Blas:2011rf} that computes the linear matter power from a given PPS. 

In these semi-numerical simulations, the resulting $21\,$cm cosmology observables are mainly sensitive to features in the matter power spectrum through their effect on the halo mass function, $\dd n(M,z)/\dd M$, that counts the number of haloes as a function of their mass. The HMF can be expressed as
\begin{align}
    \frac{\dd n}{\dd\ln M} = \frac{\overline \rho_{\rm m,0}}{M}f(\sigma)\frac{\dd \ln \sigma^{-1}}{\dd\ln M}\,,
\end{align}
where $\overline\rho_{\rm m,0}$ is the average matter density today and
\begin{align}
    \sigma^2(M,z):= D^2(z) \int \frac{{\rm d}^3 k}{(2\pi)^3} P(k) W_R^2(k)
    \label{eq:sigma_matter}
\end{align}
is the variance of the linear density field, $P(k)$, within a sphere of radius $R$ that encloses the mass $M=4\pi\overline\rho_{\rm m,0}R^3/3$. $W_R(k)$ is a window function, for which we use a top-hat function,
\begin{align}
    W_R(k)=3\frac{\sin kR-kR\cos kR}{(kR)^3}\,.
\end{align}
The multiplicity function $f(\sigma)$ usually takes a functional form with free parameters that are fit to N-body simulations. In our default scenario, we adopt the widely used Sheth--Tormen (ST) parametrisation~\cite{Sheth:1999mn} of the multiplicity function,
\begin{align}
    f_{\rm ST}(\sigma):=A\sqrt{\frac{2}{\pi}}\hat\nu\left(1+\hat\nu^{-2p}\right)e^{-\hat\nu^2/2}\,,
    \label{eq:f_st}
\end{align}
where $\hat\nu=\sqrt q\times1.686/\sigma$ and we set $A=0.353$, $p=0.175$ and $q=0.73$, as suggested by~\cite{Jenkins:2000bv}. 

However, since the formulation of Eq.~(\ref{eq:f_st}), there has been several alternative proposed parametrisations of $f(\sigma)$ to account for more recent simulations and different halo-finding methods and other modelling choices, see e.g.~\cite{Reed:2006rw,Tinker:2010my,Angulo:2012ep,Watson:2012mt,Yung:2023sib,Delos:2023eve}. Previous works have shown that the choice of HMF prescription can significantly affect the determination of both astrophysical parameters and parameters related to new physics from the $21\,$cm signal~\cite{Lopez-Honorez:2016sur, Greig:2024zso}. We therefore also investigate the impact of the assumed prescription on the determination of the running parameters. 

We compare the Sheth--Tormen fit to three alternative prescriptions: the fit of~\cite{Angulo:2012ep}, hereafter referred to as \textit{Angulo}; the redshift-dependent fit from~\cite{Watson:2012mt}, hereafter referred to as \textit{Watson}-$z$, and the more recent fit from~\cite{Yung:2023sib}, hereafter referred to as \textit{Yung}. The details on the corresponding multiplicity functions are provided in App.~\ref{app:multiplfn}. We choose these prescriptions as exemplary HMF definitions among Refs.~\cite{Jenkins:2000bv,Sheth:1999su, Reed:2006rw,Tinker:2010my,Angulo:2012ep,Watson:2012mt,Yung:2023sib,Delos:2023eve}. The Sheth--Tormen prescription serves as a historical baseline and remains the most widely used fit in $21\,$cm analyses. The Angulo prescription produces the HMFs with the largest overall amplitude in the redshift range we consider, while the Watson-$z$ produces the lowest one. Finally, Yung is the most recent proposed prescription derived in the redshift range of interest (the prescription of e.g.~\cite{Diemer:2020rgd}, while recent, was fit to N-body simulations only between $z=0.13$ and $z=2$).

\begin{figure}[t]
    \centering
    \includegraphics[width=0.85\linewidth]{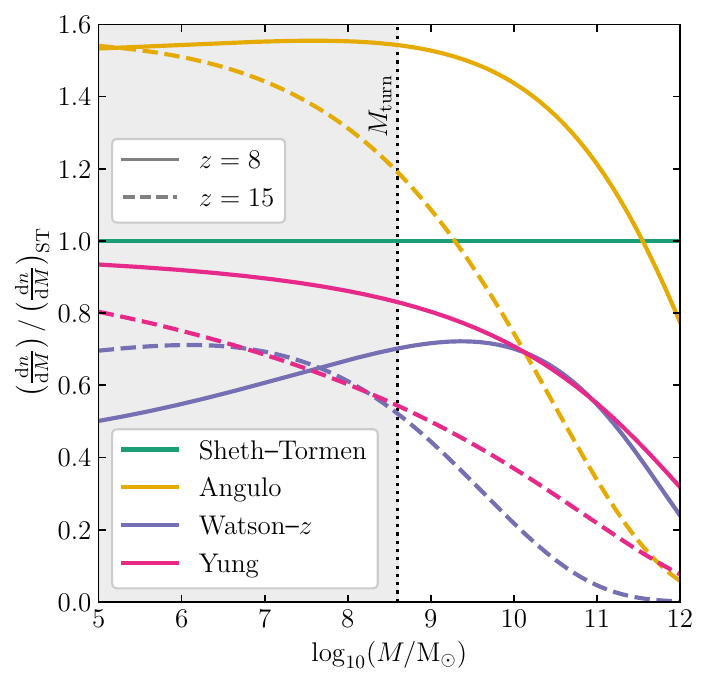}
    \caption{{\bf  Comparison between the HMF assuming different multiplicity function prescriptions.} Ratio of  Angulo (yellow), Watson-$z$ (purple) and Yung (pink) prescriptions for the HMF at $z=8$ (solid lines) and $z=15$ (dashed lines) with the Sheth--Tormen prescription (green), for fixed $\alpha_{\rm s}=\beta_\s=0$. The dotted black line shows our fiducial choice for $M_{\rm turn}$, and the grey shaded area shows the halo mass range in which star formation is exponentially suppressed.}
    \label{fig:hmf}
\end{figure}

Figure~\ref{fig:hmf} shows the ratio of the Angulo (yellow), Watson-$z$ (purple) and Yung (pink) predictions for the HMF at $z=8$ (solid lines) and $z=15$ (dashed lines) to the prediction from the Sheth--Tormen fit (green), in the mass range $10^5\leq M/{\rm M}_\odot\leq10^{12}$ and for $\alpha_\s=\beta_\s=0$. In our treatment of the $21\,$cm signal, the star formation rate is assumed to be exponentially suppressed below a given threshold mass, $M_{\rm turn}$, which is set to $10^{8.6}\,{\rm M}_\odot$ in our fiducial astrophysics model (shown as the black dotted line). Our forecasts are therefore largely insensitive to masses $M<M_{\rm turn}$, which correspond to the grey shaded area of Fig.~\ref{fig:hmf}.

\begin{figure}[t]
    \centering
    \includegraphics[width=0.99\linewidth]{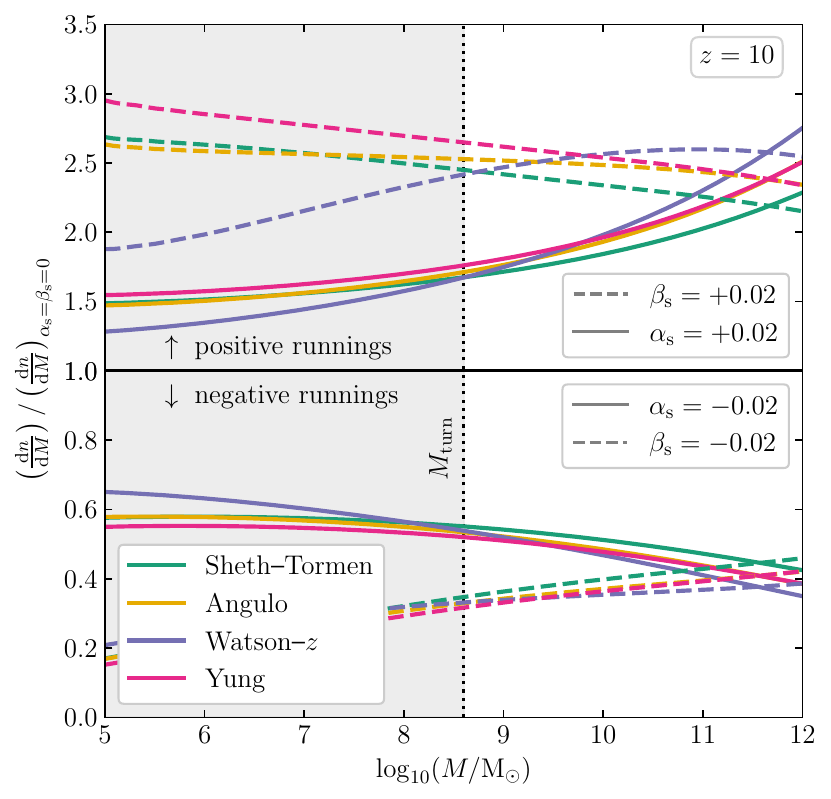}
    \caption{{\bf  Comparison between the HMF with non zero running to the one with zero running.} This plot shows the ratio of the HMF
    for $\alpha_\s=\pm0.02$ or $\beta_\s=\pm0.02$ to the HMF assuming $\alpha_\s=\beta_\s=0$ at $z=10$, as predicted by the Sheth--Tormen (green), Angulo (yellow), Watson-$z$ (purple) and Yung (pink) prescriptions respectively. The upper panel shows positive runnings, $\alpha_\s=0.02$ or $\beta_\s=0.02$, in the solid and dashed lines respectively, and the lower panel shows negative runnings, $\alpha_\s=-0.02$ or $\beta_\s=-0.02$. Positive runnings increase the halo abundance in the halo mass range of interest, negative runnings decrease it. The dotted black line shows our fiducial choice for $M_{\rm turn}$, and the grey shaded area shows the halo mass range in which star formation is exponentially suppressed.}
    \label{fig:hmf_alpha_s_beta_s}
\end{figure}

Figure~\ref{fig:hmf_alpha_s_beta_s} compares the impact of non-zero $\alpha_\s$ (solid lines) and $\beta_\s$ (dashed lines) to the $\alpha_\s=\beta_\s=0$ case for each of the four HMF prescriptions that we consider. The upper panel shows the effect of positive runnings, $\alpha_\s=+0.02$ and $\beta_\s=+0.02$, while the lower panel illustrates the effect of negative runnings, $\alpha_\s=-0.02$ and $\beta_\s=-0.02$.  We see that  both positive (negative) $\alpha_\s$ and $\beta_\s$ lead to an enhancement (suppression) of the HMF. This can be explained by considering the equivalence $M= 4\pi R^3\bar\rho_{\rm m, 0}/3$, where $k\sim 2\pi/R$. Even the largest haloes considered correspond to wavenumbers $\sim 10\,{\rm Mpc}^{-1}$, i.e. wavenumbers well above the pivot scale $k_\star$. In this $k$ range, $\alpha_\s$ and $\beta_\s$ affect the linear matter power spectrum (that seeds the HMF) qualitatively similarly, and positive (negative) $\alpha_\s$ and $\beta_\s$ lead to an enhancement (suppression) of the number of haloes with given mass $M$ w.r.t. the fiducial case.

At masses $M>M_{\rm turn}$,  we see that the Watson-$z$ HMF is the most affected by runnings (while however predicting the lowest HMF amplitude overall). On the other hand, the Sheth--Tormen HMF receives the lowest overall boost/suppression. The impact of different HMF prescriptions on the $21\,$cm signal was previously studied in~\cite{Greig:2024zso}. We  comment on the impact of the assumed HMF prescription on our results in Sec.~\ref{sec:results}.

In addition,  note that the choice of transfer function and growth factor  used in Eq.~(\ref{eq:sigma_matter}) can significantly bias the resulting output HMF. We find that using the Eisenstein \& Hu approximation~\cite{Eisenstein:1997jh} for the matter transfer function and the Liddle approximation~\cite{Liddle:1995pd} for the growth factor, which are used by default in {\tt 21cmFAST},   induce significant mass- and redshift-
dependent biases in the halo mass function with respect to the use of the outputs obtained with {\tt CLASS}. Specifically,
they lead to a non-negligible suppression of the abundance of massive haloes and consequently modify the
predicted 21 cm signal.  In our analysis we use {\tt CLASS} to set the input transfer function and growth factor for each choice of cosmological parameter set.
 We elaborate on this effect in Appendix~\ref{sec:appendix_growth_factor_transfer_function}. This discussion is
relevant for any analysis based on {\tt 21cmFAST}, beyond the modifications to the PPS considered here.

\subsection{21$\,$cm Cosmology\label{sec:21cm}}
\label{sec:21cmcosmo}

The high-redshift $21\,$cm signal from Cosmic Dawn (CD, $10\lesssim z\lesssim 30$) and the Epoch of Reionisation (EoR, around $z\sim 7$) arises from spin-flip transitions between the hyperfine states of neutral hydrogen, which is the dominant component of the intergalactic medium until reionisation. Its future measurement and detailed study is expected to open a new window on the history of the universe~\cite{Pritchard_2012}. The signal depends on the relative occupancy of the singlet and triplet states of neutral hydrogen, which is characterised by the spin temperature, $T_{\rm S}$. The observable of the high-redshift $21\,$cm signal is the differential brightness temperature, which measures its intensity relative to the CMB at the same wavelengths:
\begin{align}
    \delta T_{21}\equiv\frac{T_{\rm S}-T_{\rm CMB}}{1+z}(1-e^{-\tau_{21}})\,,
\end{align}
where $T_{\rm CMB}(z)\simeq2.73\,{\rm K}\times(1+z)$ is the CMB temperature and $\tau_{21}$ is the optical depth of the $21\,$cm line. Note that telescope arrays such as HERA or SKA are sensitive to the 21\,cm power spectrum, denoted with, $\overline{\delta T}_{21}^2(z)\Delta^2_{21}(k,z)$, where $\overline{\delta T}_{21}(z)$ is the spatial average of $\delta T_{\rm 21}$ and $\Delta_{21}^2$ refers to the dimensionless power spectrum of perturbations, which is related to the two-point correlation function of perturbations in $\delta T_{21}$ through a Fourier transform.

\subsubsection{ Astrophysics}
\label{sec:astro}

We use {\tt 21cmFAST} to perform semi-numerical simulations of the $21\,$cm signal based on the formalism introduced in~\cite{Park:2018ljd,Mesinger:2010ne}, which describes the emission of X-ray, UV and ionising radiation, which shape the signal, from astrophysical sources. The abundance of astrophysical sources depends on the HMF and thus on assumptions on the PPS. The stellar mass distribution is calculated as
\begin{align}
    \frac{\dd \rho_\star(M,z)}{\dd M}=\frac{\dd n(M,z)}{\dd M}f_{\rm duty}(M)M_\star(M)\,,
\label{eq:stmassdist}
\end{align}
where
$\rho_\star$ denotes the stellar mass density, the duty cycle $f_{\rm duty}=\exp(-M_{\rm turn}/M)$ captures the inefficiency of star formation in too light haloes, and the stellar mass of a halo with virial mass $M$ is parametrised as
\begin{align}
    M_\star(M)=M\frac{\Omega_{\rm b}}{\Omega_{\rm m}}\min\left(1,f_{\star,10}\left(\frac{M}{10^{10}\,{\rm M}_\odot}\right)^{\alpha_\star} \right)\,.
\end{align}

The X-ray emissivity depends on the stellar mass density and is defined as:
\begin{align}
    \epsilon_{\rm X}=[1+\overline\delta_{\rm b}(\textbf{x},z)]\frac{\mathcal L_{\rm X}E^{-\alpha_{\rm X}}}{t_\star H^{-1}(z)}\int_0^\infty\frac{\dd \rho_\star(M,z)}{\dd M}\dd M\,,
\end{align}
where $\overline \delta_{\rm b}$ is the average value of the nonlinear matter density contrast of the shell around $(\textbf{x},z)$, $t_\star$ is a dimensionless parameter that  characterises the timescale of star formation in units of the Hubble time, and $H(z)$ is the Hubble rate. The X-ray luminosity per star formation rate, $\mathcal L_{\rm X}$, is connected to the integrated soft-band ($E_0<E<2\,$keV) luminosity per star formation rate, $L_{\rm X}$, as
\begin{align}
    L_{\rm X}=\mathcal L_{\rm X}\int_{E_0}^{2\,\rm keV}E^{-\alpha_{\rm X}}\,\dd E\,,
\end{align}
where $E_0$ is the minimum photon energy required to escape the host halo. In our analysis, $\alpha_{\rm X}$ is fixed to its default value of 1.0.

Similarly, the number of injected ionising UV photons is given by
\begin{align}
    n_{\rm ion}(\textbf{x},z)=\frac{N_{\gamma/\rm b}}{\overline \rho_{\rm b}(\textbf{x},z)}\int_0^\infty\dd M\frac{\dd \rho_\star(M,z)}{\dd M}f_{\rm esc}(M)\,,
\end{align}
where $\overline \rho_{\rm b}(\textbf{x},z)$ is the mean baryon density in the region and the ionising UV escape fraction is modelled as
\begin{align}
    f_{\rm esc}(M)=\min\left( 1,f_{\rm esc,10} \left(\frac{M}{10^{10}\,{\rm M}_\odot}\right)^{\alpha_{\rm esc}} \right)\,.
\end{align}
$N_{\gamma/\rm b}$ is the number of ionising photons per stellar baryon that is fixed to 5000 by default.

\begin{table}[t]
    \centering
    \begin{tabular}{|c|c|c|}
    \hline
        \textbf{Parameter} & \textbf{Fiducial Value} & \textbf{Variation} \\
        \hline
        $\log_{10}f_{\star,10}$ & $-1.5$ & $\pm0.3\%$ \\
        $\alpha_\star$ & 0.5 & $\pm1\%$ \\
        $t_\star$ & 0.5 & $\pm4\%$ \\
        $\log_{10}\left(M_{\rm turn}\,/\,{\rm M}_\odot\right)$ & $8.6$ & $\pm0.4\%$ \\
        $\log_{10}f_{\rm esc,10}$ & $-1.5$ & $\pm0.25\%$ \\
        $\alpha_{\rm esc}$ & $0.0$ & $\pm0.005$ \\
        $\log_{10}\left(L_{\rm X}\,/\,\rm erg\,yr\,s^{-1}\,M_\odot^{-1}\right)$ & $40.5$ & $\pm0.04\%$ \\
        $E_0\,/\,\rm eV$ & $500$ & $\pm2.5\%$ \\
        \hline
    \end{tabular}
    \caption{Astrophysical free parameters used in section~\ref{sec:21cm}, as well as their respective fiducial values and variations used in the Fisher forecast. When shown, the $\%$ sign refers to the percentage of the fiducial value.}
    \label{tab:astro_params}
\end{table}

In total, we therefore consider eight free astrophysical parameters, listed in Tab.~\ref{tab:astro_params}. $\{f_{\star,10}, \alpha_\star, t_\star, M_{\rm turn}\}$ set the star formation history, $\{f_{\rm esc,10}, \alpha_{\rm esc}\}$ the ionization history and $\{L_\X, E_0\}$ the X-ray heating. The fiducial values were chosen in order to yield an optical depth to reionisation $\tau=0.056$  when using ST HMF, in agreement with Planck data, and closely follow the best-fit values of~\cite{Qin:2020xrg}.\footnote{Unlike~\cite{Qin:2020xrg}, we set \texttt{PHOTON\_CONS=True} in \texttt{exo21cmFAST} and weight the ionised volume fraction by the baryon density contrast in the calculation of the ionised hydrogen fraction, $\overline {x_i}=\overline{x_{\rm HII}(1+\delta_{\rm b})}$. Both changes lead to significantly larger inferred values of $\tau$, which we compensate by setting $\log_{10}f_{\star,10}=-1.5$, thus deviating from the best-fit of~\cite{Qin:2020xrg}.} Also shown in Tab.~\ref{tab:astro_params} is the variation of those parameters in the Fisher forecast, see Sec.~\ref{sec:methodology}.

\subsubsection{Imprint on the $21\,$cm signal}
\label{sec:imprint}

\begin{figure*}[t]
    \centering
    \includegraphics[width=0.49\linewidth]{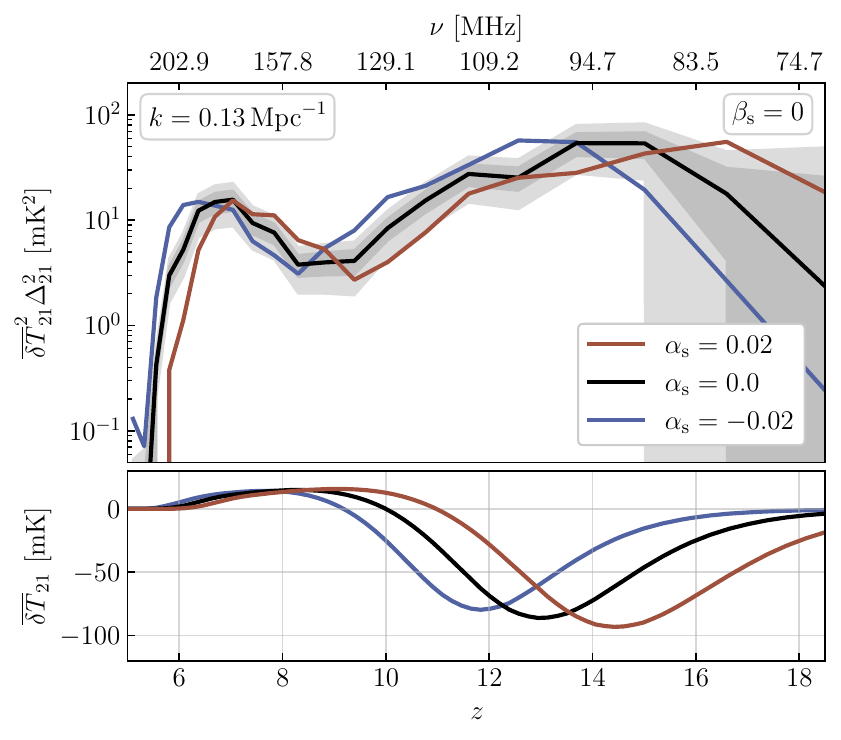}
    \includegraphics[width=0.49\linewidth]{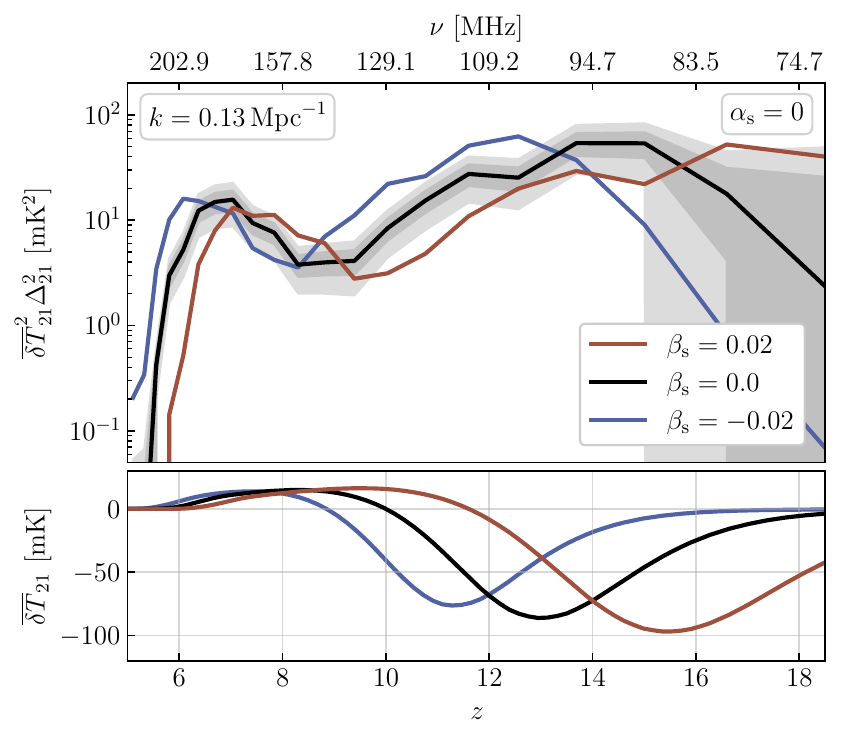}
    \caption{{\bf 21$\,$cm signal dependence on  $\alpha_\s$ and $\beta_{\rm s}$.} \textit{Upper panels}: The $21\,$cm power spectrum in the lowest $k$-bin as a function of redshift for three different choices of $\alpha_\s$ and fixed $\beta_\s=0$ (\textit{left}) and for three different choices of $\beta_\s$ and fixed $\alpha_\s=0$ (\textit{right}). The frequency of observation that each redshift corresponds to is shown at the top. The grey bands show the expected noise after 1000 hours of observation by HERA, given by Eq.~(\ref{eq:21cm_noise}), at $1\,\sigma$ and $2\,\sigma$. \textit{Lower panels:} The global (spatially averaged) $21\,$cm signal for the same scenarios. }
    \label{fig:21cm_ps_alpha_s_impact}
\end{figure*}

\begin{figure*}
    \centering
    \includegraphics[width=0.49\linewidth]{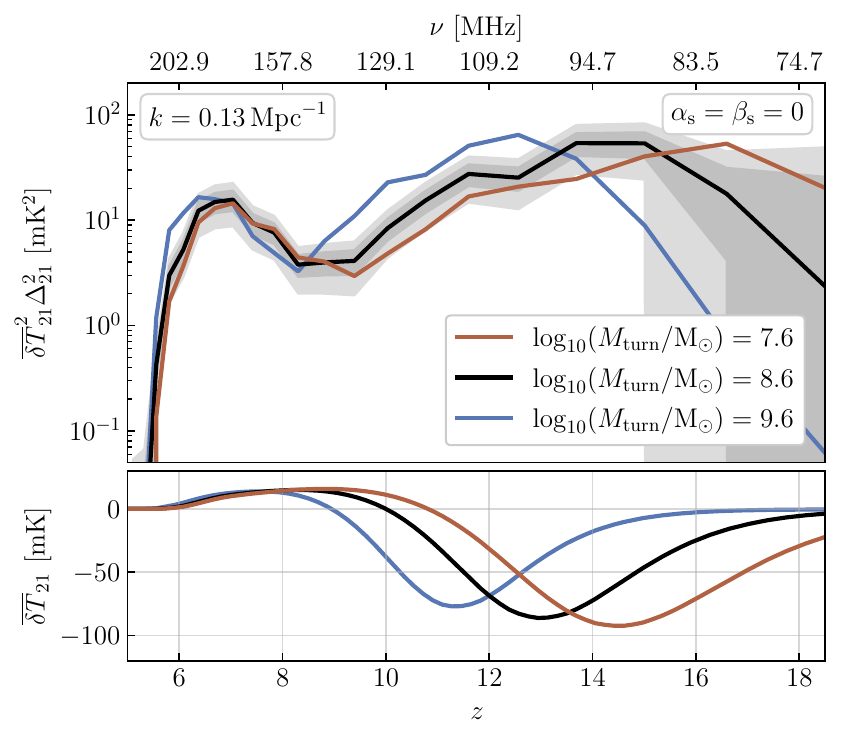}
    \includegraphics[width=0.49\linewidth]{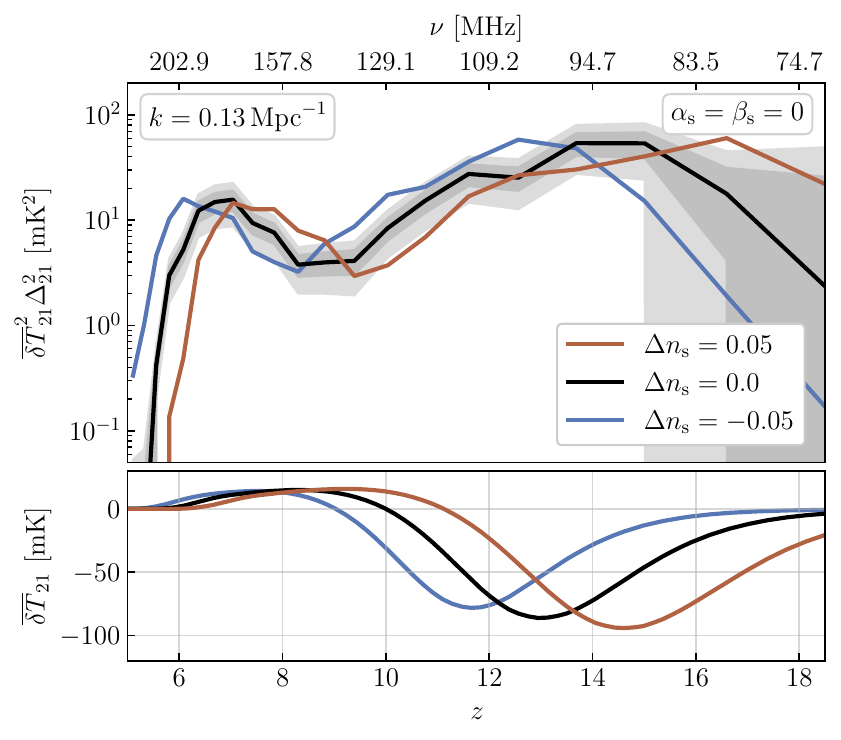}
    \caption{{\bf 21$\,$cm signal dependence on  $M_{\rm turn}$ and $n_{\rm s}$.} Same as Fig.~\ref{fig:21cm_ps_alpha_s_impact}, but showing the impact of varying $M_{\rm turn}$ (\textit{left}) and $n_{\rm s}$ (\textit{right}) on the $21\,$cm power spectrum and global signal. In the right plot, $\Delta n_{\rm s}:=n_{\rm s}-0.9665$.}
    \label{fig:ns_and_mturn_effects}
\end{figure*}

Figures~\ref{fig:21cm_ps_alpha_s_impact} and~\ref{fig:ns_and_mturn_effects} illustrate the evolution of the $21\,$cm signal as a function of redshift, and the impact thereon of a selection of model parameters. 
The lower panels show the  global (spatially averaged)  differential brightness temperature  as a function of the redshift for different scenarios. The black curve is obtained for our default set of astrophysical and cosmological parameters and zero running.
Once the first stars start to emit a significant number of Lyman-$\alpha$ photons, $T_{\rm S}$ couples to the kinetic temperature of the intergalactic medium (IGM), $T_{\rm k}<T_{\rm CMB}$, through Lyman-$\alpha$ coupling and  $\overline{\delta T}_{\rm 21}$ is seen in absorption relative to the CMB. At $z\lesssim12$ in our fiducial astrophysical model, X-ray emission from astrophysical sources begin to heat up the IGM and drive $T_{\rm S}>T_{\rm CMB}$.  $\overline{\delta T}_{\rm 21}$ is then  seen in emission relative to the CMB. Finally, starting at around $z\sim8.5$, astrophysical UV emission reionises the IGM, and the signal disappears, $\delta T_{21}\to 0$.
The upper panels show the corresponding power spectrum of perturbations, $\overline{\delta T}_{\rm 21}^2\Delta^2_{21}$, as a function of redshift inside the $k$-bin centred on $k=0.13\,{\rm Mpc}^{-1}$. Each of the three key epochs that impact $\overline{\delta T}_{21}$ described above corresponds to a peak in the power spectrum, owing to the spatial inhomogeneities in the processes that cause them. The grey bands indicate the expected measurement sensitivity of HERA given $1000$ hours of observation. For more details, see Sec.~\ref{sec:methodology}.

The effect of variations of $\alpha_\s$ (left panel) and $\beta_\s$ (right panel) on the $21\,$cm power spectrum and global signal is shown in Fig.~\ref{fig:21cm_ps_alpha_s_impact} with coloured lines. It can be seen that non-zero positive (negative) values of those parameters lead to features appearing at earlier (later) times in both the global signal and the power spectrum. This can be understood as positive $\alpha_\s$ and $\beta_\s$ both increase the number of haloes of a given mass, implying more and earlier star formation and thus stronger and earlier Lyman-$\alpha$, X-ray and ionising fluxes, shifting the signal forward in time. Note that we do not show $\alpha_\s,\beta_\s=\pm0.05$ like we did in Figs.~\ref{fig:pk_alphas_betas} and~\ref{fig:cmb_tcl}, because those values would dramatically  impact the $21\,$cm power spectrum in Fig.~\ref{fig:21cm_ps_alpha_s_impact}. 

The effect of $\alpha_\s$ and $\beta_\s$ is expected to be at least partially degenerate with the effects of variations of other astrophysical and cosmological parameters. Fig.~\ref{fig:ns_and_mturn_effects} shows the impact of variations of $M_{\rm turn}$ (\textit{left}) and $n_{\rm s}$ (\textit{right}) on the observables. As in Fig.~\ref{fig:21cm_ps_alpha_s_impact}, more yellow-brownish colours correspond to larger parameter values and bluer colours to lower values. Decreasing $M_{\rm turn}$, the threshold mass for star formation allows increasingly small haloes, which are more abundant than larger haloes, to contribute to star formation and the emission of Lyman-$\alpha$ and X-ray photons, thereby leading to earlier Lyman-$\alpha$ coupling and X-ray heating. This shifts the global $21\,$cm signal to earlier times, similarly to the effect of positive running parameters. The timing of reionisation is however largely unaffected (for this choice of fiducial model). On the other hand, an increase in $n_\s$ effectively increases the amplitude of the PPS at $k>k_\star$ and therefore also the amplitude of the HMF. This leads to more haloes of all masses and a shift of the $21\,$cm signal to earlier times, including reionisation.

As we will see in Sec.~\ref{sec:results}, these are  two parameters that exhibit the strongest degeneracies with $\alpha_{\rm s}$.

\section{\label{sec:methodology_datasets}Datasets}

In our analysis in Sec.~\ref{sec:results}, we use the following datasets.

\subsubsection{Planck Data}

Using TTTEEE+lowE+lensing data from the 2018 data release, the Planck collaboration set constraints of~\cite{Planck:2018jri}:
\begin{align}
    \alpha_\s=-0.0045\pm0.0067
\end{align}
when fixing $\beta_\s=\Delta N_{\rm eff}=0$ and
\begin{eqnarray}
    \alpha_\s&=&0.002\pm0.010 \\
    \beta_\s&=&0.010\pm0.013
\end{eqnarray}
when varying $\alpha_\s$ and $\beta_\s$, but fixing $\Delta N_{\rm eff}=0$. In order to evaluate Planck constraints on the model with free $\alpha_\s$, $\beta_\s$ and $\Delta N_{\rm eff}$, we perform a separate MCMC on the same data (but with the inclusion of BAO measurements from~\cite{Beutler:2011hx,Ross:2014qpa,BOSS:2016wmc}). We use the MCMC sampler \texttt{Cobaya}\footnote{\url{https://github.com/CobayaSampler/cobaya}}~\cite{Torrado:2020dgo,2019ascl.soft10019T}, which is interfaced with \texttt{CAMB}\footnote{\url{https://github.com/cmbant/CAMB}}~\cite{Lewis:1999bs,Howlett:2012mh}. We marginalise over all cosmological parameters given in Tab.~\ref{tab:cosmo_parameters} as well as foreground nuisance parameters. Non-linear corrections to the matter power spectrum, which the lensing measurements are sensitive to, are computed using {\tt HMcode~2020}~\cite{Mead:2015yca,Mead:2020vgs}. {\tt HMcode} might not be able to accurately capture the effects of $\alpha_\s$ or $\beta_\s$; however, the corrections provided by {\tt HMcode} to constraints from lensing are already small and we therefore expect this effect to be negligible. We check convergence using the Gelman-Rubin criterion~\cite{gelman1992inference}, requiring that $R-1<0.01$. We find
\begin{eqnarray}
    \alpha_\s&=&-0.002\pm0.011 \\
    \beta_\s&=&0.007\pm0.013 \\
    \Delta N_{\rm eff}&=&-0.13\pm0.20\,.
\end{eqnarray}
In this case, $\Delta N_{\rm eff}$ is only weakly correlated with $\alpha_\s$ or $\beta_\s$. We discuss the sensitivity of Planck measurements to $\alpha_\s$, $\beta_\s$ and $\Delta N_{\rm eff}$ further in Sec.~\ref{sec:results}.

\subsubsection{Simons Observatory Mock Data}

The ground-based Simons Observatory~\cite{SimonsObservatory:2018koc,SimonsObservatory:2025wwn} (SO) will measure the large-$\ell$ tail of the CMB temperature and polarisation spectra with great accuracy. To evaluate combined constraints from Planck data and SO mock data, we perform a separate MCMC, using the publicly available \texttt{MFlike}\footnote{\url{https://github.com/simonsobs/LAT_MFLike}} likelihood code for the TTTEEE spectra and the \texttt{SOLikeT}\footnote{\url{https://github.com/simonsobs/soliket}} likelihood for the SO lensing measurements, both of which are interfaced with {\tt Cobaya}. In the combined analysis, we disable Planck lensing inference and exclusively use mock SO lensing measurements. We restrict Planck to scales $\ell\leq1000$ for the TT spectrum and $\ell\leq600$ for the TE and EE spectra, and use SO mock data for $\ell>1000$ and $\ell>600$ respectively.

\subsubsection{HERA Mock Data}

We perform a Fisher forecast to assess HERA's upcoming ability to constrain the inflationary power spectrum, using 1000 hours of observations of the high-redshift $21\,$cm signal from Cosmic Dawn and the Epoch of Reionisation. The simulations are performed using the public code \texttt{exo21cmFAST}, based on \texttt{21cmFAST}. We perform the Fisher forecasts using \texttt{21cmCAST}\footnote{\url{https://github.com/gaetanfacchinetti/21cmCAST}}~\cite{Facchinetti:2023slb}.

We assume an astrophysical scenario dominated by a single population of galaxies that host mostly Population II stars, with fiducial astrophysical parameters given in Tab.~\ref{tab:astro_params}. We set {\tt PHOTON\_CONS}$=${\tt True} for photon conservation corrections~\cite{Park:2021eux} in order to derive meaningful ionization histories. The photon conservation algorithm is not yet implemented into {\tt 21cmFAST} in the presence of Population III stars hosted in minihaloes, and, as a result, a self-consistent computation of $\tau$ would not be possible.

In our analysis, we compute the $21\,$cm power spectrum in bins of $k$ and $z$, chosen in size such that the separation of the bin centres corresponds to a bandwidth of $8\,$MHz. This follows the convention of~\cite{Facchinetti:2023slb,Mason:2022obt} and allows us to assume that the noise in each of the bins is uncorrelated. This leads to 15 $k$-bins between $0.1<k/{\rm Mpc}^{-1}<1$, which is the range least affected by foregrounds and thermal noise, and 21 redshift bins between $z=5$ and $z=22$. In Figs.~\ref{fig:21cm_ps_alpha_s_impact} and~\ref{fig:ns_and_mturn_effects}, we show the lowest $k$-bin, centred on $k=0.13\,{\rm Mpc}^{-1}$, in which HERA is expected to be the most sensitive to the $21\,$cm power spectrum.

The expected noise from 1000 hours of observation by HERA, projected on the fiducial power spectrum of Figs.~\ref{fig:21cm_ps_alpha_s_impact}  and~\ref{fig:ns_and_mturn_effects} with grey bands at $1\,\sigma$ and $2\,\sigma$,  has the following contributions:
\begin{align}
    \sigma_{21}^2 = \sigma_{\rm exp}^2 + \sigma_{\rm shot}^2 + \left[f_{\rm noise} \times \overline{\delta T}_{\rm 21}^2\Delta^2_{21}\right]^2\,.
    \label{eq:21cm_noise}
\end{align}
We calculate the experimental error $\sigma_{\rm exp}$ using the Python package \texttt{21cmSense}\footnote{\url{https://github.com/jpober/21cmSense}}~\cite{Pober:2012zz,Pober:2013jna} (assuming a frequency channel width of $100\,$kHz and the {\tt moderate} foreground model). On the other hand, $\sigma_{\rm shot}$ accounts for Poisson (shot) noise in each bin and the choice of $f_{\rm noise}=0.2$ is meant to account for modelling uncertainties in the $21\,$cm power spectrum~\cite{Park:2018ljd}. We note that forecast results for $21\,$cm cosmology may  depend on the exact choice of $f_{\rm noise}$~\cite{Dandoy:2025twl}, which we will discuss further in Sec.~\ref{sec:results}.

\section{Methodology}
\label{sec:methodology}

When the posterior of a model with parameters $\bm\theta$ given some data $\bm d$ can be assumed to be a multivariate Gaussian, one may approximate the covariance matrix of the model posterior as the inverse of the Fisher matrix
\begin{align}
    F_{ij} \equiv -\mathbb{E} \left[ \left. \frac{\partial^2 }{\partial \theta_i\, \partial \theta_j} \ln \mathcal{L}(\bm d \, |\, \bm\theta)  \, \right| \, \bm\theta \right]  \, ,
\end{align}
where $\mathcal{L}(\bm d\,|\,\bm\theta)$ is the likelihood of the data given the model and the expectation value $\mathbb{E} [..] $ means the average over all possible data realizations generated by the model with parameters $ \bm\theta$. For the HERA forecast, we assume a diagonal covariance
\begin{align}
    F_{ij}=\sum_{i_k}\sum_{i_z}\left[\frac{1}{\sigma_{21}^2}\frac{\partial\, \overline{\delta T}_{\rm 21}^2\Delta^2_{21}}{\partial\, \theta_i} \frac{\partial\, \overline{\delta T}_{\rm 21}^2\Delta^2_{21}}{\partial\, \theta_j}\right]_{({i_k},{i_z})}\,,
\end{align}
where $i_k$ and $i_z$ run over all $k$- and $z$-bins.

The derivatives are evaluated using the finite difference method, for which we compute the $21\,$cm power spectrum at the fiducial point in parameter space (which is automatically the best-fit) as well as for small variations of each parameter $\theta_i^{\rm fid}\pm\delta_{\theta_i}$. If this variation is chosen too small, numerical noise dominates the derivative, whereas if it is chosen too large, the response of the observable to the variation becomes nonlinear. For each parameter $\theta_i$, we adjust $\delta_{\theta_i}$ such that $\Delta\chi^2\sim1$,\footnote{
$\Delta\chi^2=-2 (\ln {\cal L} (\theta_i^{\rm fid}+\delta_{\theta_i})-\ln {\cal L} (\theta^{\rm fid}_i))$, where ${\cal L}$ is a Gaussian likelihood with variance $\sigma^2_{21}$.} 
which produces forecast uncertainties approximately independent of $\delta_{\theta_i}$~\cite{Brinckmann:2018cvx}.\footnote{We have checked that this method produces similar results as the method used in~\cite{Dandoy:2025twl}, where the target observable was computed on a grid of values of $\delta_{\theta_i}$ and then a plateau away from numerical noise was identified in which $\sigma_{\theta_i}$ was independent of $\delta_{\theta_i}$.} The chosen parameter variations for the HERA forecast are shown in Tabs.~\ref{tab:cosmo_parameters} and~\ref{tab:astro_params}.

To evaluate combined constraints between the mock 21\,cm observations from HERA and the CMB inference from Planck data or the Planck+SO combination, we extract the covariance matrices from the CMB MCMCs, which describe the Gaussian posterior. In theory, the combined constraints could then be evaluated as $C^{-1}_{\rm combined}=C^{-1}_{\rm HERA}+C^{-1}_{\rm CMB}$. However, in the $21\,$cm forecast, $\tau$ is not an input parameter of \texttt{exo21cmFAST}, but is instead forward-modelled from the input parameters $\bm\theta$ as $\tau=f(\bm\theta)$, meaning that a priori it does not appear in $C_{\rm HERA}$ and a simple inverse addition of the covariance matrix would miss the constraining power that the very accurate determination of $\tau$ from 21\,cm observations adds to the determination of cosmological parameters from CMB observations. To calculate the expected uncertainty on $\tau$ and its correlations with other parameters, we propagate the errors on $\bm\theta$ by linearizing around the fiducial parameter values $\bm\theta_0$ as $\tau=\tau_0+\bm J\times (\bm\theta-\bm\theta_0)$, where $\tau_0=f(\bm\theta_0)$ and $\bm J=\dd f/\dd\bm\theta$ is the Jacobian of $\tau$ with respect to the model parameters. The covariance matrix $C_{\rm HERA}=F^{-1}_{\rm HERA}$ for the model parameters $\bm \theta$ can thus be enlarged to the parameters space $( {\bm \theta}, \tau)$ as
\begin{align}
    C^{(\tau)}_{\rm HERA}=\begin{pmatrix}
        C_{\rm HERA} & C_{\rm HERA}\bm J \\
        (C_{\rm HERA}\bm J)^{\rm T} & \bm J^{\rm T}C_{\rm HERA}\bm J
    \end{pmatrix}\,. \label{eq:appended_covmat}
\end{align}
Although convenient for display purposes, the resulting matrix is rank-deficient and singular and $F^{(\tau)}_{\rm HERA}= (C^{(\tau)}_{\rm HERA})^{-1}$ does not exist. We can therefore not evaluate combined CMB and HERA constraints by adding the corresponding Fisher matrices in the enlarged parameter space. To evaluate such combined constraints, we impose the relation $\tau=f(\bm\theta)$ on the CMB Fisher matrix before adding it to $F_{\rm HERA}$. This amounts to a projection the CMB constraints on the subspace of independent parameters $\bm \theta$,
\begin{align}
 F_{\rm CMB}^{({\bm \theta)}}  = \begin{pmatrix}
        \mathbb{1}_{\rm dim(\bm \theta)} \\
        \bm J
    \end{pmatrix}^{\rm T}F_{\rm CMB}\begin{pmatrix}
        \mathbb{1}_{\rm dim(\bm \theta)} \\
        \bm J
    \end{pmatrix}\,,
\end{align}
which can then be combined with the HERA Fisher matrix as $F_{\rm tot}^{({\bm \theta})}= F_{\rm HERA}+F_{\rm CMB}^{({\bm \theta})}  $. We then reconstruct the uncertainty on $\tau$ using Eq.~(\ref{eq:appended_covmat}).

Given a Fisher matrix, the separation between a point in parameter space $\tilde{\bm\theta}$ and the fiducial $\tilde{\bm\theta}_0$ is quantified by the Mahalanobis distance
\begin{align}
    D=\sqrt{(\tilde{\bm\theta}-\tilde{\bm\theta}_0)^{\rm T}\tilde C^{-1}(\tilde{\bm\theta}-\tilde{\bm\theta}_0)}\,.
\end{align}
Here, $\tilde{\bm\theta}$ and has been restricted to only those parameters whose value differs from their fiducial values, and $\tilde C$ is the corresponding subblock of $C$. This corresponds to a detection at
\begin{align}
    Z=\Phi^{-1}\left(1-P(\chi^2_{\dim \tilde{\bm\theta}}\geq D^2)/2\right)
\end{align}
two-sided Gaussian-equivalent sigmas, where $\Phi$ is the cumulative distribution function of 1 one-dimensional Gaussian and $\chi^2_{\dim \tilde{\bm\theta}}$ is the chi-squared distribution with $\dim\tilde{\bm\theta}$ degrees of freedom.

\section{\label{sec:results}Results}

In Sec.~\ref{sec:cons_as}, we present our forecasts on the model consisting of the astrophysical parameters of Tab.~\ref{tab:astro_params}, six cosmological base parameters, and $\alpha_\s$, which we denote as
\begin{eqnarray}
    {\bm \theta}_{\alpha_s}&=&\{\log_{10}f_{\star,10}, \alpha_\star, t_\star, \log_{10}\left(M_{\rm turn}\,/\,{\rm M}_\odot\right), \log_{10}f_{\rm esc,10}, \nonumber\\
     &&\alpha_{\rm esc}, \log_{10}\left(L_{\rm X}\,/\,\rm erg\,yr\,s^{-1}\,M_\odot^{-1}\right),E_0\,/\,\rm keV\nonumber\\
     &&\omega_{\rm b}, \omega_{\rm cdm}, h, \ln(10^{10} A_\s), n_\s, \tau, \alpha_\s
     \}\, .
     \label{eq:param_as}
\end{eqnarray}
Here, $\beta_\s$ and $\Delta N_{\rm eff}$ remain fixed to their fiducial values of Tab.~\ref{tab:cosmo_parameters}. In Sec.~\ref{sec:cons_bs}, we present our forecasts on the extended parameter space including $\beta_\s$:
\begin{equation}
    {\bm \theta}_{\beta_s}=\{  {\bm \theta}_{\alpha_s},\beta_s\}
  \label{eq:param_bs}\,.
\end{equation}
Then, in Sec.~\ref{sec:cons_Neff}, we present our forecasts on the full parameter space
\begin{equation}
    {\bm \theta}_{\Delta N_{\rm eff}}=\{  {\bm \theta}_{\beta_s},\Delta N_{\rm eff}\}
  \label{eq:param_Neff}\,,
\end{equation}
to compare to~\cite{Garny:2026gcs}. Finally, we discuss the impact of the choice of HMF prescription and assumptions on the error in Secs.~\ref{sec:cons_HMF} and~\ref{sec:cons_sig} respectively.

We summarise our main results in Tab.~\ref{tab:forecast_summary}.

\begin{table*}
    \begin{tabular}{|l||c||c|c||c|c|c|}
        \hline
        \textbf{Model} & $\bm\theta_{\alpha_\s}$ & \multicolumn{2}{c||}{$\bm\theta_{\beta_\s}$} & \multicolumn{3}{c|}{$\bm\theta_{\Delta N_{\rm eff}}$} \\
        \hline
        \textbf{Parameter} & $\alpha_\s$ & $\alpha_\s$ & $\beta_\s$ & $\alpha_\s$ & $\beta_\s$ & $\Delta N_{\rm eff}$ \\
         \hline
         \hline
        \textbf{HERA} & $\pm$0.070 & $\pm$0.081 & $\pm$0.037 & $\pm$0.084 & $\pm$0.039 & $\pm$1.4 \\
        \textbf{Planck} & $\pm$0.0066 & $\pm$0.010 & $\pm$0.012 & $\pm$0.011 & $\pm$0.013 & $\pm$0.20 \\
        \textbf{Planck+SO} & $\pm$0.0040 & $\pm$0.0041 & $\pm$0.0073 & $\pm$0.0067 & $\pm$0.0082 & $\pm$0.12 \\
        \textbf{HERA+Planck} & $\pm$0.0060 & $\pm$0.0074 & $\pm$0.0088 & $\pm$0.0077 & $\pm$0.0088 & $\pm$0.065 \\
        \textbf{HERA+Planck+SO} & $\pm$0.0038 & $\pm$0.0039 & $\pm$0.0068 & $\pm$0.0046 & $\pm$0.0069 & $\pm$0.051 \\
        \hline
    \end{tabular}
    \caption{All resulting forecast uncertainties on the parameters $\{\alpha_\s,\beta_\s,\Delta N_{\rm eff}\}$ in the $\bm\theta_{\alpha_\s}$, $\bm\theta_{\beta_\s}$ and $\bm\theta_{\Delta N_{\rm eff}}$ models. We show all combinations of experiments that we consider: HERA, Planck, Planck+SO as well as HERA+Planck and HERA+Planck+SO.}
    \label{tab:forecast_summary}
\end{table*}

\subsection{Constraints on $\alpha_\s$}
\label{sec:cons_as}
\begin{figure}[t]
    \centering
    \includegraphics[width=\linewidth]{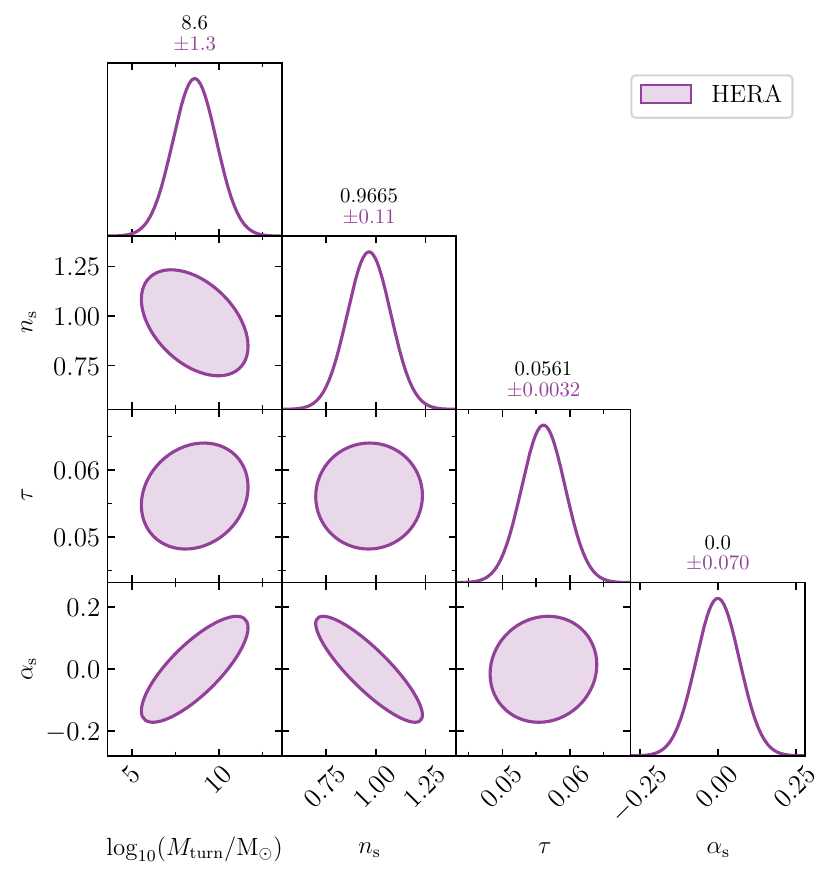}
    \caption{{\bf HERA Fisher forecast for the ${\bm \theta}_{\alpha_\s}$ model}. We show the marginalised 2D (off-diagonal plots) and 1D (diagonal plots) confidence intervals of a selection of parameters. 2D contours are shown at the $95\%$ confidence limit. Fiducial values as well as the marginalised 1D confidence intervals for each parameter at $68\%$ CL are given above the diagonal plots. All other parameters not shown in this plot but considered in our analysis have been marginalised over. A full version of this plot showing all considered parameters can be found in Fig.~\ref{fig:alpha_s_large}.}
    \label{fig:alpha_s_hera_planck}
\end{figure}

\begin{figure}[t]
    \centering
    \includegraphics[width=\linewidth]{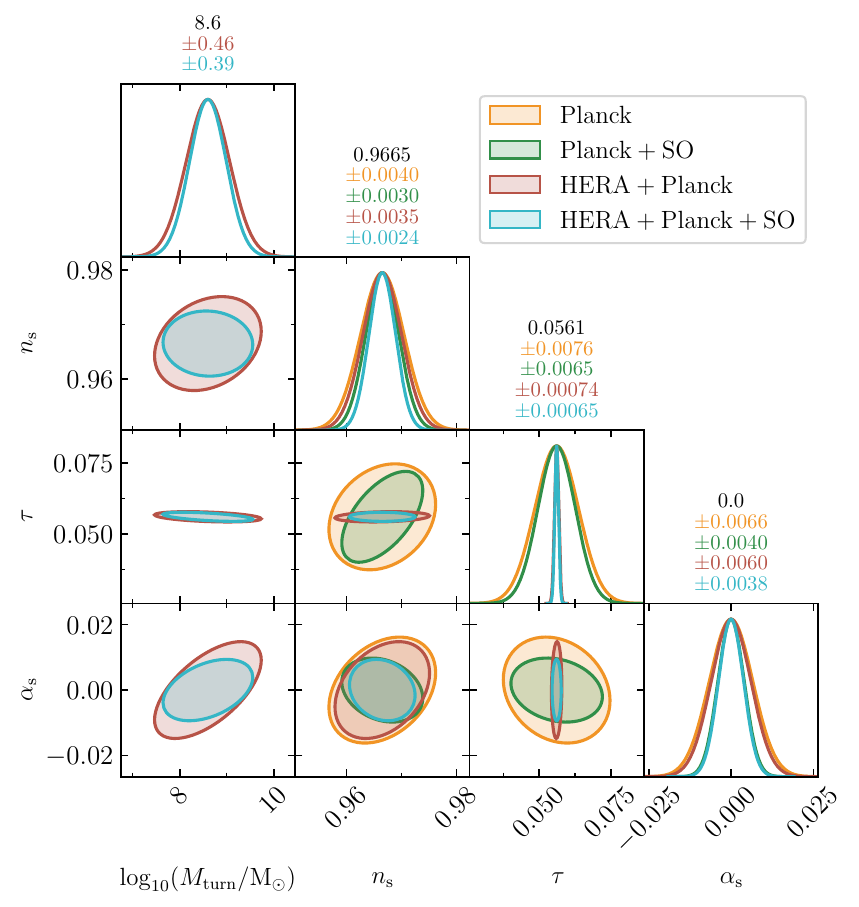}
    \caption{{\bf Planck + SO + HERA forecasts for the ${\bm \theta}_{\alpha_s}$ model}. Same as Fig.~\ref{fig:alpha_s_hera_planck}, but for the data combinations of Planck (orange), Planck+SO (green), HERA+Planck (terracotta) and HERA+Planck+SO (blue). A full version of this plot is contained in Fig.~\ref{fig:alpha_s_large}.}
    \label{fig:alpha_s_so}
\end{figure}

We first discuss our forecasts on ${\bm \theta_{\alpha_s}}$. Fig.~\ref{fig:alpha_s_hera_planck} shows the resulting constraints on a subset of the model parameter space, namely $\{M_{\rm turn}, n_{\rm s}, \tau, \alpha_\s\}$, from mock HERA observations. Fig.~\ref{fig:alpha_s_so} combines our HERA forecasts with current and future CMB bounds. A full version of these plots with all considered parameters can be found in Fig.~\ref{fig:alpha_s_large} in the appendix. As can be seen in the full plot, while CMB observatories dominate constraints on $\omega_{\rm cdm}, \omega_{\rm b}, h, A_\s$ and $n_\s$, the 21\,cm experiment is able to constrain $\tau$ much more accurately as it is not subject to the same $A_\s$--$\tau$ degeneracy~\cite{Clesse:2012th,Liu:2015gaa,Shmueli:2023box}. With no external prior on $\Lambda$CDM parameters from CMB observations, the uncertainty on the inferred value of $\tau$ from high-redshift $21\,$cm observations, $\sigma_\tau^{\rm HERA}=0.0032$, is already much lower than that from Planck data, $\sigma_\tau^{\rm Planck}=0.0076$. As the determination of $\tau$ from $21\,$cm observations (anti-) correlates significantly with several cosmological parameters, most notably $\omega_{\rm b}$, $\omega_{\rm cdm}$ and $h$, all of which are very well-constrained by the CMB, the combination of HERA and Planck data leads to a ten-fold improvement, $\sigma_\tau^{\rm HERA+Planck}=0.00074$. This in turn leads to a better determination of $\Lambda$CDM parameters from Planck, most notably $A_\s$ and $\omega_{\rm cdm}$.

On its own, HERA does not provide an improvement on the determination of $\alpha_\s$ (when $\beta_\s$ is fixed to zero) over existing CMB bounds. This may be attributed to the poor sensitivity of HERA towards all $\Lambda$CDM parameters (except $\tau$) as well as the strong degeneracy between $\alpha_\s$ and $n_\s$, both of which influence the amplitude of the halo mass function on small scales and lead to very similar features in the $21\,$cm signal, see Figs.~\ref{fig:21cm_ps_alpha_s_impact} and~\ref{fig:ns_and_mturn_effects}. To a slightly lesser extent, there also exists a degeneracy between $\alpha_\s$ and $M_{\rm turn}$, which impacts the $21\,$cm signal similarly at $z\gtrsim9$, but not at lower redshifts, leading to a milder degeneracy (compare Figs.~\ref{fig:21cm_ps_alpha_s_impact} and~\ref{fig:ns_and_mturn_effects} as well as the discussion in Sec.~\ref{sec:21cmcosmo}). Because Planck is able to constrain $n_\s$ very precisely, combining the two datasets breaks the $\alpha_\s$-$n_\s$ degeneracy in HERA data.

These degeneracies may be understood by comparing with Fig.~\ref{fig:hmf_alpha_s_beta_s}: non-zero values of $\alpha_\s$ change the abundance of heavier haloes more strongly than they change the abundance of lighter haloes with $M\sim M_{\rm turn}$. This mimics the effect of $n_\s$, which also primarily affects the abundance of heavy haloes. $M_{\rm turn}$ controls how strongly lighter haloes ($10^8\lesssim M/{\rm M}_\odot\lesssim10^9$) contribute to the $21\,$cm signal and could therefore somewhat counteract the change of abundance of lighter haloes induced by non-zero values of $\alpha_\s$, but not the change of heavier haloes, leading to a different (and weaker) kind of degeneracy.

Figure~\ref{fig:alpha_s_so} shows the constraints on $\{M_{\rm turn},\tau,n_\s,\alpha_\s\}$ that result from Planck data (orange) as well as the Planck+SO (green), HERA+Planck (terracotta) and HERA+Planck+SO (blue) combinations. Including HERA data with Planck data leads to only a slightly reduced uncertainty, $\sigma_{\alpha_\s}^{\rm Planck}=0.0066$ to $\sigma_{\alpha_\s}^{\rm HERA+Planck}=0.0060$, which is still weaker than the $\sigma_{\alpha_\s}^{\rm Planck+ACT+BAO}=0.0052$ result from~\cite{AtacamaCosmologyTelescope:2025nti}. The Simons Observatory will be able to access larger multipoles/smaller scales than Planck, which leads to a greater sensitivity to $\alpha_\s$. In this case, we forecast $\sigma_{\alpha_\s}^{\rm Planck+SO}=0.0040$, which is a significant improvement over Planck+ACT. However, adding HERA data does not further improve sensitivity to $\alpha_\s$. In addition to $\tau$ and $A_\s$, constraints on $\omega_{\rm cdm}$ and $h$ do nevertheless significantly benefit from this data combination (compare Fig.~\ref{fig:alpha_s_large}).

A model with $\alpha_\s=-0.01$ was suggested in~\cite{Rogers:2023upm} as one possible solution to a tension between the amplitude and tilt of the matter power spectrum as inferred from CMB and Lyman-$\alpha$ forest measurements. Note that this  preference for non-zero $\alpha_\s$ (or non-zero $\alpha_\s$ and $\beta_\s$) disappears when using a more recent likelihood which adopts more conservative modelling of astrophysics and uncertainties~\cite{Fairbairn:2025fko}. Nevertheless, from the analysis presented here, it would be possible to detect such a model at $2.5\,\sigma$ ($2.6\,\sigma$) significance with the Planck+SO (HERA+Planck+SO) data combinations.

\subsection{Constraints on $\alpha_\s$ and $\beta_\s$}
\label{sec:cons_bs}

\begin{figure}[t]
    \centering
    \includegraphics[width=\linewidth]{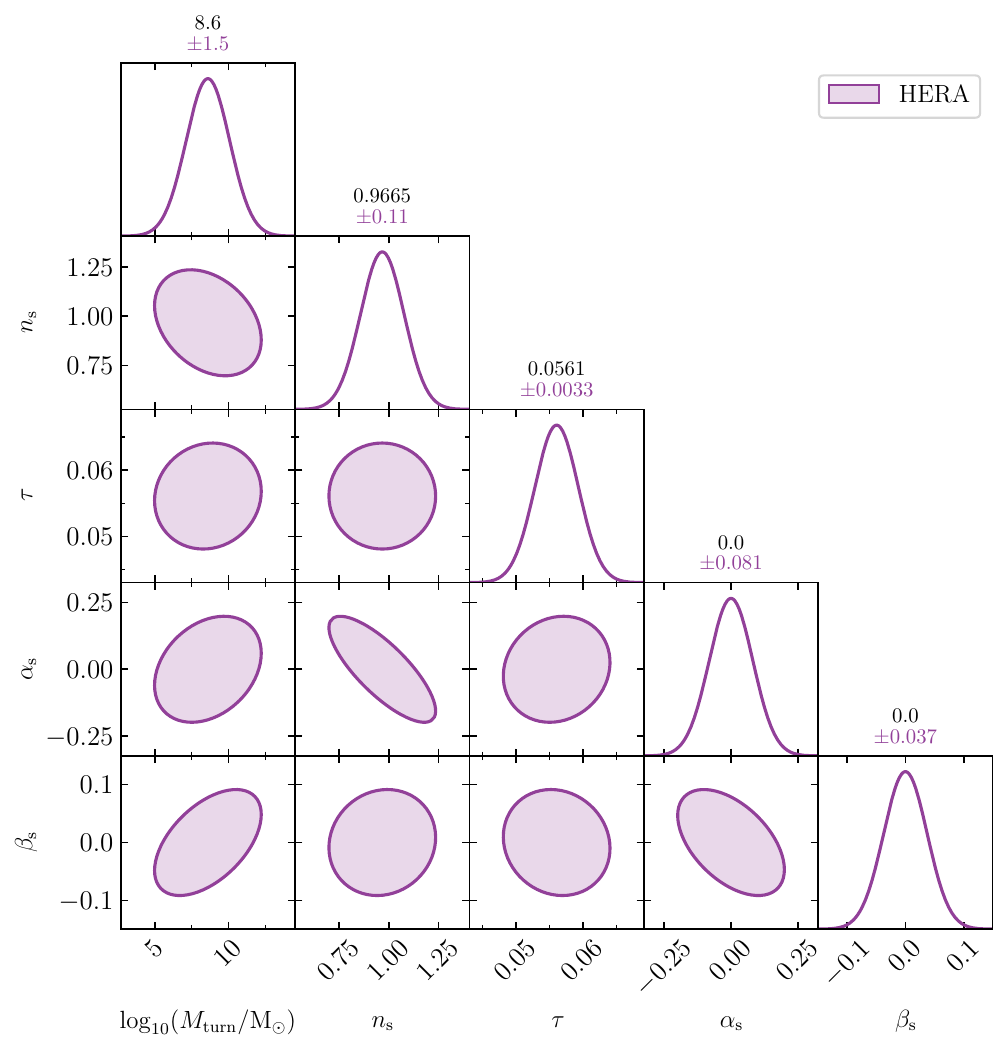}
    \caption{{\bf HERA Fisher forecast for the ${\bm \theta}_{\beta_s}$ model}. Same as Fig.~\ref{fig:alpha_s_hera_planck}, but considering ${\bm \theta}_{\beta_s}$.}
    \label{fig:beta_s_hera_planck}
\end{figure}
\begin{figure}[t]
    \centering
    \includegraphics[width=\linewidth]{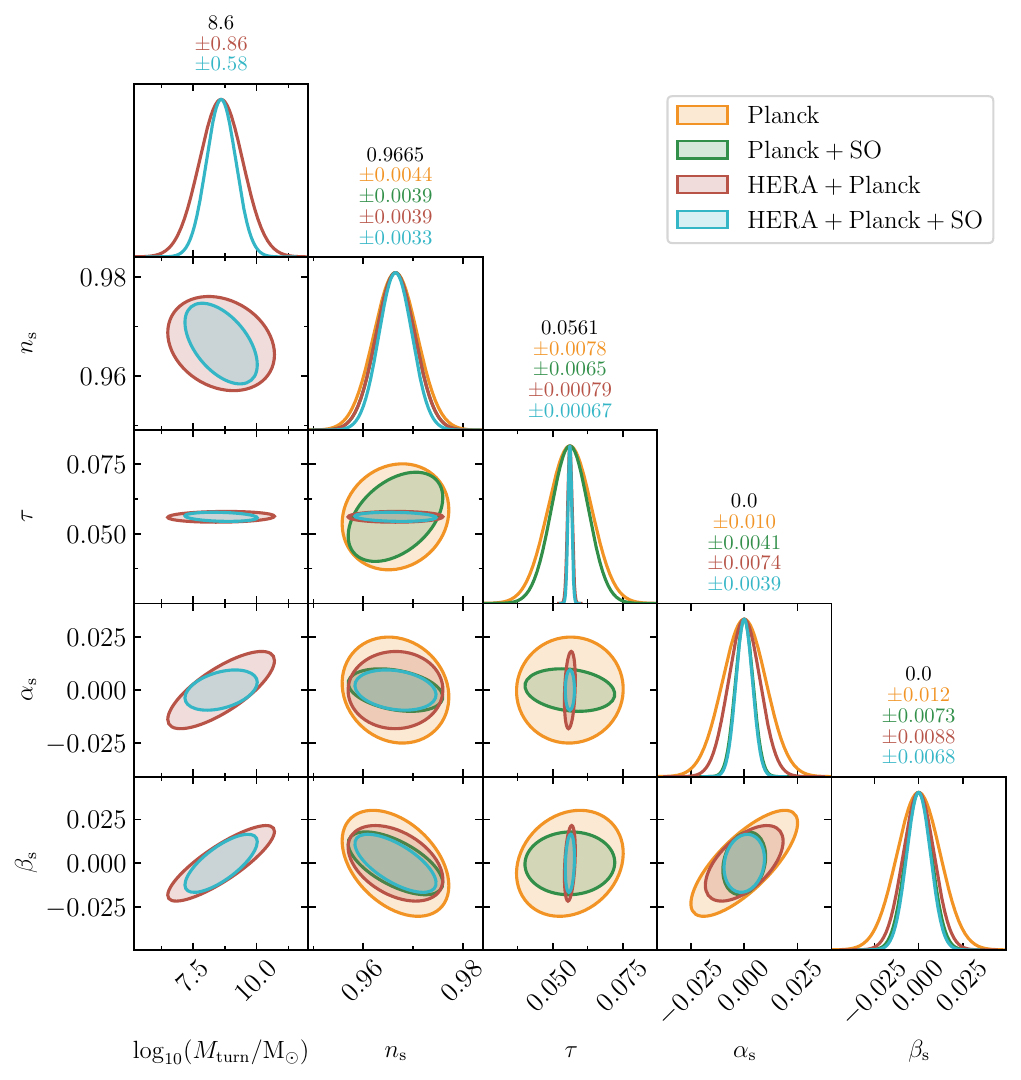}
    \caption{{\bf Planck + SO + HERA forecasts for the ${\bm \theta}_{\beta_s}$ model}. Same as Fig.~\ref{fig:alpha_s_so}, but for ${\bm \theta}_{\beta_s}$.}
    \label{fig:beta_s_so}
\end{figure}

We now consider the extended set of parameters ${\bm \theta}_{\beta_s}$. Figs.~\ref{fig:beta_s_hera_planck} and~\ref{fig:beta_s_so} show the same as Figs.~\ref{fig:alpha_s_hera_planck} and~\ref{fig:alpha_s_so} respectively, except that now $\beta_\s$ is varied in addition to $\alpha_\s$, rather than being fixed to zero, as it was before. A full plot showing all parameters can be found in Fig.~\ref{fig:beta_s_large}.

Where $\alpha_\s$ exhibited a significant degeneracy with $n_\s$ in HERA mock data, $\beta_\s$ is most degenerate with $M_{\rm turn}$. This can again be understood by comparing with Fig.~\ref{fig:hmf_alpha_s_beta_s}: for both non-zero values of $\alpha_\s$ and changes in $n_\s$, the abundance of heavy haloes is the most strongly affected, leading to a degeneracy. Non-zero values of $\beta_\s$ however act more strongly on the abundance of lighter haloes (with the exception of the Watson-$z$ HMF, shown in purple in Fig.~\ref{fig:hmf_alpha_s_beta_s}, for positive values of $\beta_\s$). The contribution of lighter haloes with $M\sim M_{\rm turn}$ to the $21\,$cm signal is controlled by $M_{\rm turn}$, meaning that an increase (decrease) of $M_{\rm turn}$ values could counteract the effect of an increased (decreased) $\beta_\s$ more strongly than it can counteract the effect of an increased (decreased) $\alpha_\s$. This leads to the degeneracies (and lack thereof) between $\alpha_\s$, $\beta_\s$, $n_\s$ and $M_{\rm turn}$ that can be observed in Fig.~\ref{fig:beta_s_hera_planck}.

It is also worth noting that because positive (negative) values of $\alpha_\s$ and $\beta_\s$ both lead to an enhancement (suppression) of the HMF in the mass range relevant for HERA (which corresponds to $k>k_\star$), they are significantly anti-correlated in HERA data. In Planck data however, the two parameters are correlated, which may be understood as meaning that Planck constraints on $\alpha_\s$ and $\beta_\s$ mainly come from $k<k_\star$ range, where positive (negative) $\beta_\s$ affects the amplitude of the PPS in the opposite direction to positive (negative) $\alpha_\s$, compare Figs.~\ref{fig:pk_alphas_betas} and~\ref{fig:cmb_tcl}. We have checked that if one were to consider just SO mock data alone, $\alpha_\s$ and $\beta_\s$ would be anti-correlated.

Compared to the ${\bm\theta}_{\alpha_s}$ model, where the inclusion of HERA data brought a $\sim10\%$ improvement in constraining power to $\alpha_\s$ over Planck data alone, there is a greater improvement in constraining power in the case of the ${\bm\theta}_{\beta_s}$ model. Constraints on $\alpha_\s$ and $\beta_\s$ improve by $\sim25\%$ each in the Planck+HERA data combination, compared to just Planck alone, and the degeneracy that was present between $\alpha_\s$ and $\beta_\s$ in Planck data gets alleviated (but not resolved).

In this model, the inclusion of SO data leads to a $\sim60\%$ reduction of the uncertainty on $\alpha_\s$ from $\pm0.010$ to $\pm0.0041$ over just Planck data alone. This is significantly tighter than the $\pm0.0074$ uncertainty that HERA+Planck achieve, and the inclusion of HERA data to Planck+SO data does not lead to a further improvement in constraining power (within the forecast uncertainty). Constraints on $\beta_\s$ from Planck+SO ($\pm0.0073$) are also better than those from HERA+Planck ($\pm0.0088$). In this case, the inclusion of HERA data leads to a modest further $\sim7\%$ reduction on the uncertainty on $\beta_\s$ w.r.t. Planck+SO alone.

\subsection{Constraints on $\alpha_\s$, $\beta_\s$ and $N_{\rm eff}$}
\label{sec:cons_Neff}

It has been pointed out~\cite{Garny:2026gcs} that there exists a significant degeneracy in CMB observations between $\Delta N_{\rm eff}$ and $\{\alpha_\s, \beta_\s\}$. Constraints on either $\Delta N_{\rm eff}$ or $\{\alpha_\s, \beta_\s\}$ thus strongly depend on whether the other quantities are allowed to vary or kept fixed to zero, as in the base $\Lambda$CDM model. It is therefore interesting to test whether the degeneracy could be lifted by combining CMB data with $21\,$cm data from HERA. Non-zero values of $\Delta N_{\rm eff}$ can be constrained from $21\,$cm observations through the parameter's effect on the matter power spectrum, as has been done in~\cite{Plombat:2024kla}. Concretely, a positive value of $\Delta N_{\rm eff}$ leads to a suppression of the matter power spectrum on all scales that the $21\,$cm signal is sensitive to, and a negative value to an enhancement. The authors of~\cite{Plombat:2024kla} found that while combining HERA data with that of Planck and SO does lead to a modest improvement in constraints on $\Delta N_{\rm eff}$ over Planck+SO data alone, the effect is small.\footnote{Our inferred uncertainty on $\Delta N_{\rm eff}$ from HERA alone when fixing $\alpha_\s$ and $\beta_\s$ to zero is worse than that of~\cite{Plombat:2024kla} ($\sigma_{N_{\rm eff}}=1.3$ compared to $0.84$), while our uncertainty inferred from the combination of HERA with Planck and SO data is significantly better than theirs ($\sigma_{N_{\rm eff}}=0.043$ compared to $0.090$). This discrepancy may be due to any number of reasons related to the simulation setup. One possibility is that our setup excludes minihaloes/Pop III stars, which \cite{Plombat:2024kla} found to be a source of degeneracy with $\Delta N_{\rm eff}$, and includes CMB lensing.} However, this was done for $\alpha_\s=\beta_\s=0$; when deviations of the power-law primordial power spectrum are allowed, CMB constraints on $\Delta N_{\rm eff}$ relax significantly. It is therefore interesting to test to what extent 21\,cm observations could be used to break this degeneracy.
h
\begin{figure}[t]
    \centering
    \includegraphics[width=0.8\linewidth]{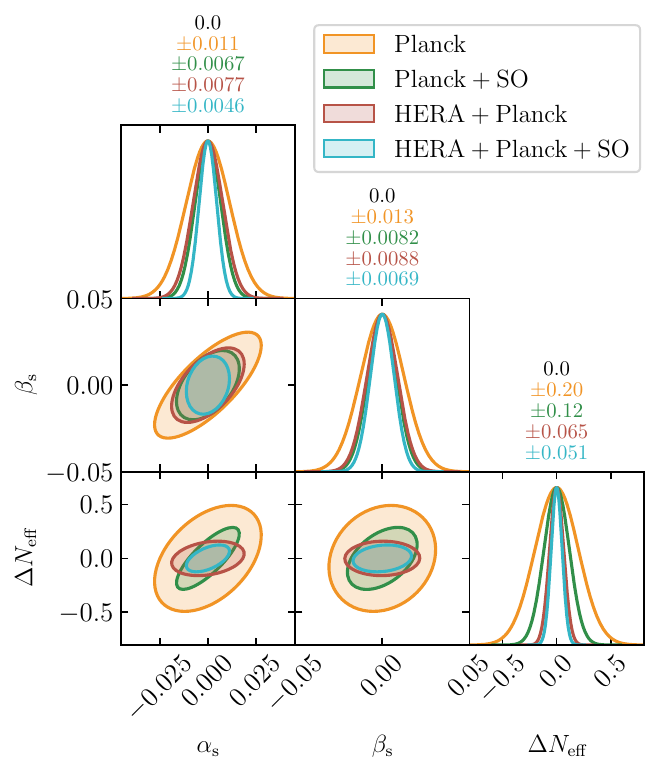}
    \caption{{\bf Impact of including $\Delta N_{\rm eff}$.} Same as Fig.~\ref{fig:alpha_s_so}, considering $\theta_{\Delta N_{\rm eff}}$.}
    \label{fig:neff}
\end{figure}

Figure~\ref{fig:neff} shows the resulting constraints on the $\bm\theta_{\Delta N_{\rm eff}}$ model from Planck (orange), Planck+SO (green), HERA+Planck (terracotta) and HERA+Planck+SO (blue). In Planck data, again $\alpha_\s$ and $\beta_\s$ are correlated. Because $\Delta N_{\rm eff}$ affects the high-$\ell$ tail of the CMB power spectra, there exists a slight degeneracy between $\Delta N_{\rm eff}$ and $\alpha_\s$, but not $\Delta N_{\rm eff}$ and $\beta_\s$, as constraints on the latter mostly come from low-$\ell$ measurements when considering just Planck data. Again, the inclusion of SO mock data results in a significant reduction in the uncertainties on $\alpha_\s$ and $\beta_\s$. However, with $\Delta N_{\rm eff}$ now free, bounds on $\alpha_s$ relax by over $60\%$ from $\pm0.0041$ to $\pm0.0067$ in the Planck+SO case, and the uncertainty on $\Delta N_{\rm eff}$ relaxes from $\pm0.071$ in the case of $\alpha_\s=\beta_\s=0$ (not pictured) to $\pm0.12$ when both parameters are allowed to vary, confirming the degeneracy pointed out in Ref.~\cite{Garny:2026gcs}. There is also a smaller, $12\%$ relaxation of constraints on $\beta_\s$ from $\pm0.0073$ to $\pm0.0082$.

One important distinction between the work of~\cite{Garny:2026gcs} and ours however is that we assume positive $\Delta N_{\rm eff}$ to translate into a non-zero abundance of a free-streaming dark radiation (DR) species, whereas~\cite{Garny:2026gcs} considered a model of self-interacting DR, which is less constrained by CMB data than standard free-streaming DR. They furthermore considered an additional model where the self-interacting DR is initially coupled to dark matter and undergoes a DR-dark matter decoupling at a redshift $z_{\rm dec}$, which is an additional free parameter in the theory. The latter model relaxed constraints on all cosmological parameters, including $h$ and $\Delta N_{\rm eff}$ slightly more than the baseline self-interacting DR model did. The degeneracy between $\Delta N_{\rm eff}$ and $\{\alpha_\s,\beta_\s\}$ is therefore expected to be stronger in their model than in ours, however extending our analysis to such a model would be outside the scope of this work.

Interestingly, we find that HERA+Planck have a (slightly) lower sensitivity to $\alpha_\s$ and $\beta_\s$ than Planck+SO do ($\sigma_{\alpha_\s}^{\rm HERA+Planck}=0.0077$ and $\sigma_{\beta_\s}^{\rm HERA+Planck}=0.0088$ compared to $\sigma_{\alpha_\s}^{\rm Planck+SO}=0.0067$ and $\sigma_{\alpha_\s}^{\rm HERA+Planck}=0.0082$), but are significantly more constraining to $\Delta N_{\rm eff}$ ($\sigma_{\Delta N_{\rm eff}}^{\rm HERA+Planck}=0.065$ compared to $\sigma_{\Delta N_{\rm eff}}^{\rm Planck+SO}=0.12$). The full combination of HERA, Planck and SO (mock) data leads constraints at $\sigma_{\alpha_\s}^{\rm HERA+Planck+SO}=0.0046$ ($30\%$ better than Planck+SO alone) and $\sigma_{\Delta N_{\rm eff}}^{\rm HERA+Planck+SO}=0.051$ (almost $60\%$ better than Planck+SO alone). This once again highlights the interesting synergies between 21\,cm observations from Cosmic Dawn and the Epoch of Reionisation and next-stage CMB measurements, in this case for disentangling features in the primordial power spectrum from the effect of dark radiation species.

By combining only subsets of the relevant covariance matrices, we have explicitly checked that the constraining power that HERA adds over Planck and the SO alone w.r.t. $\{\alpha_\s,\beta_\s,\Delta N_{\rm eff}\}$ is neither simply a consequence of the more accurate determination of $\tau$ by HERA, nor the more accurate determination of $\{\omega_{\rm cdm}, \omega_{\rm b}, h, \ln(10^{10}\,A_\s), n_\s\}$ from CMB experiments. Instead, the combined constraining power is really a consequence of synergies between the full combination of the two datasets.

A model with $\alpha_\s=\beta_\s=0.02$ could be detected with the HERA+Planck+SO data combination at a significance of $5.3\,\sigma$ with $\Delta N_{\rm eff}$ fixed to $0$, or at $4.4\,\sigma$ when $\Delta N_{\rm eff}$ is allowed to vary. Planck+SO alone could make an almost equally strong $4.8\,\sigma$ detection of such a model when $\Delta N_{\rm eff}$ is assumed to be fixed to zero, however when allowing $\Delta N_{\rm eff}$ to vary, the detection significance decreases to $2.8\,\sigma$. When considering just HERA+Planck data, the same model could be detected at $2.4\,\sigma$ almost independent of whether $\Delta N_{\rm eff}$ remains fixed or is allowed to vary.

A model with $\alpha_\s=\beta_\s=0.02$ and $\Delta N_{\rm eff}=0.5$ could be detected at $7.5\,\sigma$ and $9.5\,\sigma$ by HERA+Planck and HERA+Planck+SO respectively. A free-streaming analogue of the best-fit point from~\cite{Garny:2026gcs} could thus be tested at high significance; a dedicated analysis of the SIDR models of~\cite{Garny:2026gcs} is left for future work.

\subsection{Effect of HMF Prescription}
\label{sec:results_hmf_effect}
\label{sec:cons_HMF}

\begin{figure}[t]
    \centering
    \includegraphics[width=\linewidth]{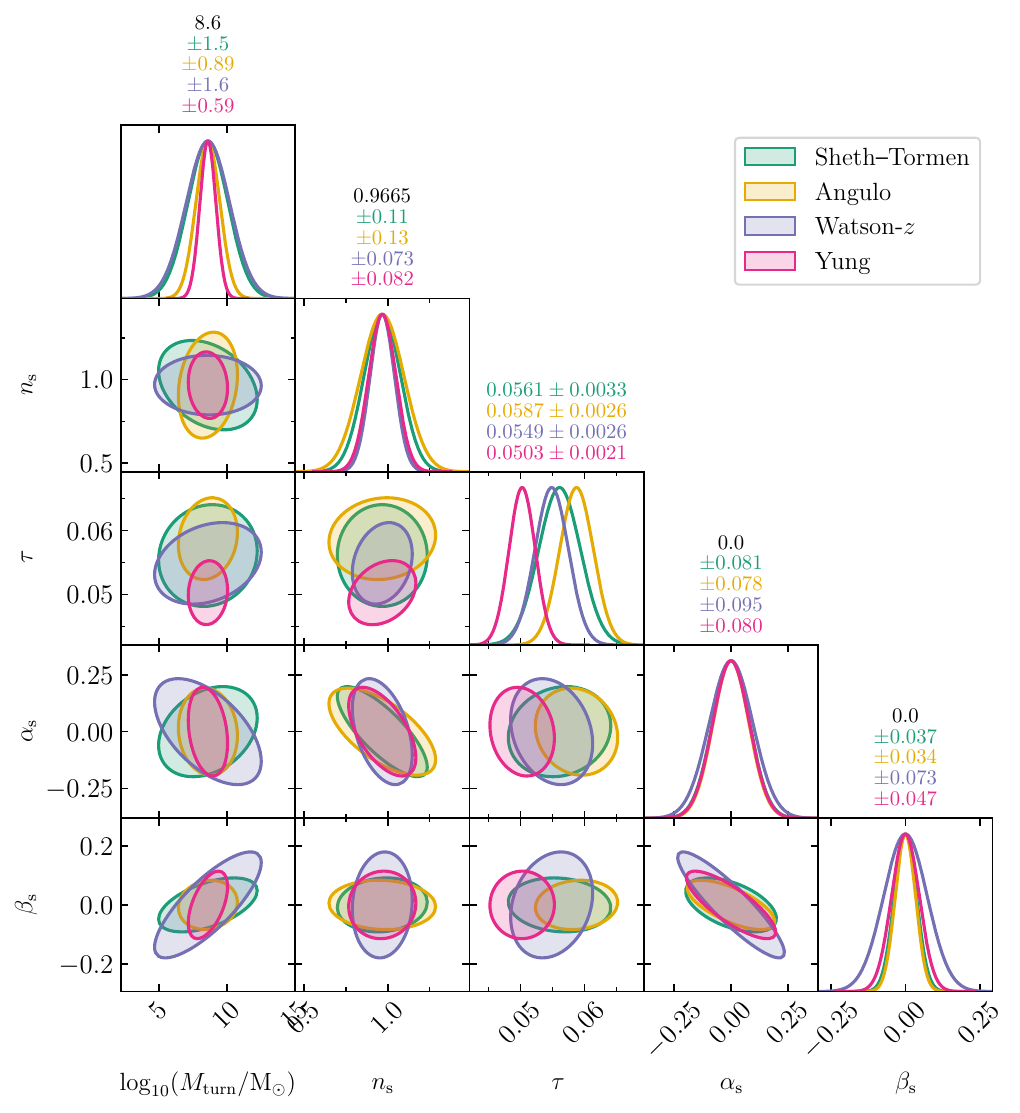}
    \caption{{\bf Impact of the HMF prescription.} Same as Fig.~\ref{fig:beta_s_hera_planck}, showing the resulting posterior from HERA alone on the ${\bm \theta}_{\beta_\s}$ model, when using either the Sheth--Tormen (green), Angulo (yellow), Watson-$z$ (purple) or Yung (pink) HMF prescriptions. Because $\tau$ is a derived quantity here, its central value changes with the HMF prescription.}
    \label{fig:posterior_hmf_comparison}
\end{figure}

As discussed in Sec.~\ref{sec:HMF},  the choice of HMF
prescription in simulation codes like {\tt 21cmFAST} is known to affect the determination of model parameters from 21\,cm cosmology experiments. This is illustrated in Fig.~\ref{fig:posterior_hmf_comparison} for the ${\bm\theta_{\beta_s}}$ set of parameters. The plot compares the resulting posteriors of the model from HERA alone under the assumption of the Sheth--Tormen HMF (green), Angulo (yellow), Watson-$z$ (purple) and Yung (pink) HMF prescriptions. The most significant differences arise in the parameter posteriors of $M_{\rm turn}$ and $\beta_s$. The error on $M_{\rm turn}$ differs by more than a factor $2.5$ between the Yung and Watson-$z$ parametrisations, and by almost a factor of two between the Angulo and Watson-$z$ prescriptions.

Because $\tau$ is a derived quantity in this model, both its central value and its error change with the HMF prescription. All obtained central values are however still within $1\,\sigma$ uncertainty of the central value for Planck optical depth. 

The differences between the different posteriors greatly decrease when the HERA mock data is combined with Planck and SO data. The uncertainty between posteriors on cosmological parameters agrees to within $\sim20\%$ (the worst case is $\beta_\s$), and the uncertainties on $M_{\rm turn}$ agree within $\sim50\%$. The combination of the relevant covariance matrices is not strictly allowed as the central values of $\tau$ do not match; we therefore do not show a plot for illustration. Instead we evaluate the shift, or bias, $\delta\theta_\alpha$, w.r.t. the fiducial parameter values of $n_\s,\alpha_\s,\beta_\s$ and $M_{\rm turn}$ reported in Tabs.~\ref{tab:cosmo_parameters} and~\ref{tab:astro_params} when performing parameter inference on the mock data assuming a HMF model that differs from the "true". The bias on a parameter $\theta_\alpha$ is estimated as 
~\cite{Linder:2006xb}
\begin{align}\delta\theta_\alpha=C_{\alpha\beta}\sum_{i_k}\sum_{i_z}\left[\frac{\partial\mathcal O_{\rm t}}{\partial\theta_{\beta}}\frac{\mathcal O_{\rm t}-\mathcal O_{\rm w}}{\sigma^2_{\rm t}}\right]\,.
    \label{eq:linder_bias}
\end{align}
 where $\mathcal O_{\rm t}$  and $\mathcal O_{\rm w}$ refer to the relevant observable computed assuming the "true" or "wrong" HMF model.  In Tab.~\ref{tab:hmf_posterior_biases} we report those biases using HERA+Planck+SO data and assuming  the Sheth-Tormen HMF as the "true" model. Most biases are relatively small except for $\beta_\s$ in the Angulo and Yung cases where it reaches more than two standard deviations.

\begin{table}[ht]
    \centering
    \begin{tabular}{|c||c|c|c|c|}
    \hline
        \multirow[c]{2}{*}[-0.35em]{\textbf{HMF Model}} & \multicolumn{4}{c|}{\bf Parameter Bias (HERA+Planck+SO)} \\
        \cline{2-5}
        & $\log_{10}\left(\frac{M_{\rm turn}}{{\rm M}_\odot}\right)$ & $n_\s$ & $\alpha_\s$ & $\beta_\s$ \\
        \hline
        Angulo & $0.48$ & $0.00031$ & $-0.0035$ & $-0.0098$ \\
        Watson & $-0.11$ & $0.00021$ & $-0.00044$ & $-0.0014$ \\
        Yung & $0.57$ & $0.0029$ & $-0.0021$ & $-0.016$ \\
        \hline
    \end{tabular}
    \caption{Biases on select parameters when inferred assuming the wrong HMF parametrisation (here Angulo, Watson or Yung in the first columns) in parameter inference from mock data generated assuming the Sheth-Tormen HMF, as calculated using Eq.~(\ref{eq:linder_bias}).}
    \label{tab:hmf_posterior_biases}
\end{table}

\subsection{Effect of the Noise Fraction}
\label{sec:cons_sig}

Through Eq.~(\ref{eq:21cm_noise}), our constraints depend on the fraction of the value of the $21\,$cm power spectrum added to the total noise as modelling error, denoted as $f_{\rm noise}$. We adopt a fiducial value of $f_{\rm noise}=0.2$, i.e. $20\%$ modelling error, following the convention of~\cite{Park:2018ljd}. In Tab.~\ref{tab:noise_fraction_uncertainties}, we show the effect of different values of the noise fraction, namely $f_{\rm noise}=\{0.3, 0.2, 0.1\}$, on our obtained constraints on the $\bm\theta_{\Delta N_{\rm eff}}$ model in the HERA+Planck+SO data combination. The uncertainty of $\alpha_\s$ does not exhibit a strong dependence on the noise fraction. The uncertainty of $\beta_\s$ almost does not change between $f_{\rm noise}=0.3$ and $f_{\rm noise}=0.2$, but does decrease by $\sim10\%$ from $f_{\rm noise}=0.2$ to $f_{\rm noise}=0.1$. This could be interepreted such that constraining power on $\beta_\s$ is not limited by noise, but rather by its degeneracy with $M_{\rm turn}$. The most strongly affected uncertainty is that on $\Delta N_{\rm eff}$, which decreases from $0.060$ to $0.051$ to $0.038$ from $f_{\rm noise}=0.3$ to $0.2$ to $0.1$ respectively. This is because HERA dominates the constraints on $\Delta N_{\rm eff}$ in this model, making the inferred constraining power sensitive to assumptions on the noise.

\begin{table}[h]
    \centering
    \begin{tabular}{|c|c c c|}
    \hline
        \textbf{Noise fraction} & $\sigma_{\alpha_\s}$ & $\sigma_{\beta_\s}$ & $\sigma_{\Delta N_{\rm eff}}$ \\
        \hline
        $f_{\rm noise}=0.3$ & $0.0048$ & $0.0070$ & $0.060$ \\
        $f_{\rm noise}=0.2$ & $0.0046$ & $0.0069$ & $0.051$ \\
        $f_{\rm noise}=0.1$ & $0.0042$ & $0.0064$ & $0.038$ \\
        \hline
    \end{tabular}
    \caption{Uncertainties on $\{\alpha_\s,\beta_\s,\Delta N_{\rm eff}\}$ in the combination of HERA, Planck and SO data as a function of the noise fraction, $f_{\rm noise}$, in Eq.~(\ref{eq:21cm_noise}).}
    \label{tab:noise_fraction_uncertainties}
\end{table}

\subsection{HERA and SKA AA4 Comparison}

We calculate the expected noise of SKA observations using the AA4 configuration of SKA in {\tt 12cmSense}, but otherwise identical noise assumptions as for HERA.

\begin{figure}[t]
    \centering
    \includegraphics[width=0.99\linewidth]{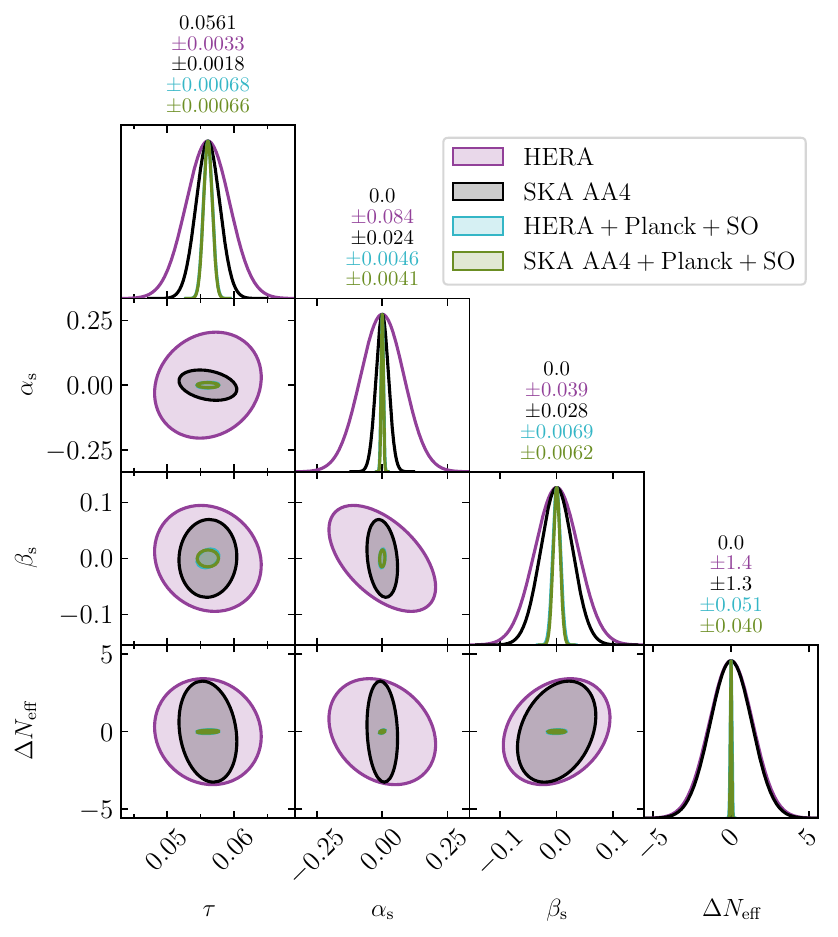}
    \caption{{\bf Comparison between HERA and SKA forecasts for the $\bm\theta_{\Delta N{\rm eff}}$ model.}}
    \label{fig:hera_ska_comparison}
\end{figure}

Figure~\ref{fig:hera_ska_comparison} compares the posteriors of select parameters of the $\bm\theta_{\Delta N{\rm eff}}$ model. On its own, SKA achieves a $45\%$ improvement in constraining power on $\tau$, a $70\%$ improvement on $\alpha_\s$ and a $30\%$ improvement on $\beta_\s$,\footnote{This order of magnitude agrees with the difference in constraining power between HERA and SKA that previous works have found on different set of parameters, such as in~\cite{Agius:2025nfz} for decaying dark matter, in~\cite{Facchinetti:2026wed} for primordial magnetic fields, and in~\cite{Agius:2025xbj} for primordial black holes (as discussed in private communication with the authors of that work).} but no improvement in constraining power on $\Delta N_{\rm eff}$, as the limiting factor in this case is the almost perfect degeneracy between $\Delta N_{\rm eff}$ and $A_\s$ in the large-$k$ linear matter power spectrum. In combination with Planck and SO however, constraints on $\tau$ and $\alpha_\s$ only improve marginally over HERA+Planck+SO. SKA achieves a $15\%$ improvement on the uncertainty on $\beta_\s$ and a more significant $25\%$ improvement on the constraining power towards $\Delta N_{\rm eff}$.

\section{\label{sec:conclusion}Conclusion}

Future observations of the high-redshift 21\,cm signal are expected to provide a powerful probe of the history of the universe during Cosmic Dawn and the Epoch of Reionisation. In this work, we have investigated the sensitivity of HERA and SKA to departures from a power-law primordial power spectrum, focusing on the running of the scalar spectral index, $\alpha_\s$, and the running thereof, $\beta_\s$. We have also examined the complementarity of these measurements with present and future CMB observations. 

In our analysis based on \texttt{exo21cmFAST} simulations, the 21\,cm signal is  primarily sensitive to the impact of runnings of the primordial power spectrum through their impact on the halo mass function. We show that positive (negative) values of both $\alpha_\s$ and $\beta_\s$ enhance (decrease) the abundance the haloes above the threshold mass for star forming galaxies, leading to an earlier (later) onset of Lyman-$\alpha$ coupling, X-ray heating, and reionisation, all of which are processes that shape the 21\,cm signal.

Considering a minimal set of cosmological and astrophysics parameters with the addition of $\alpha_\s$, our Fisher matrix analysis shows that HERA alone does not outperform current Planck constraints because of strong degeneracies, particularly with $n_\s$ and, to a lesser extent, with the astrophysical parameter $M_{\rm turn}$. However, combining HERA with Planck partially breaks these degeneracies and allows to marginally improve the constraint on $\alpha_\s$ by about 8\%. When both $\alpha_\s$ and $\beta_\s$ are varied simultaneously, the complementarity becomes more pronounced: the combination of HERA and Planck improves constraints on both parameters by roughly 25\% relative to Planck alone while partially reducing the degeneracy between them. The inclusion of Simons Observatory (SO) mock data further strengthens the sensitivity to primordial spectral runnings through its access to smaller scales in the CMB. In this case, HERA does not bring additional constraining power to $\alpha_\s$ over Planck combined with the SO alone, although it does improve constraints on reionisation-related quantities and helps in reducing residual degeneracies, particularly in models including $\beta_\s$. 

CMB bounds on $\alpha_\s$ and $\beta_\s$ rely strongly on assumptions on the effective number of fermionic relativistic degrees of freedom, $N_{\rm eff}$, and CMB bounds on $N_{\rm eff}$ in turn rely strongly on assumptions on the primordial power spectrum. We have shown that HERA is able to break this degeneracy and lead to significantly tighter constraints on those parameters than obtained from Planck alone or Planck combined with the Simons Observatory. The improvement is the strongest for $N_{\rm eff}$, for which the inclusion of HERA mock data leads to more than a factor two better determination than with Planck and SO alone.

A major strength of 21\,cm cosmology is its ability to constrain the reionisation history and the optical depth. It is already well known that HERA can determine $\tau$ significantly more precisely than Planck does. Combining 21\,cm and CMB observations will therefore substantially improve constraints on several standard cosmological parameters through the breaking of parameter degeneracies, most notably between $\tau$ and $A_\s$. However, we checked that the strong combined constraining power of HERA in combination with CMB experiments is not just a result of the better determination of $\tau$ from 21\,cm observations, but really an outcome of  synergies between the two types of datasets.

We have also assessed the robustness of our results to modelling assumptions. Different halo mass function prescriptions are known to induce biases in the reconstruction of the model parameters. Here we have checked that the uncertainties on parameters that only HERA is sensitive to can improve (worsen) by a factor up to $\sim 2$ depending on the chosen prescription. However, these discrepancies get drastically reduced when combining the 21\,cm mock observations with CMB data, and the resulting combined uncertainties on cosmological parameters all agree within $\sim20\%$. On the other hand, varying the assumed level of 21\,cm modelling uncertainty modifies the forecasts only moderately for most parameters, indicating that our main conclusions are not driven by specific choices in the analysis. Finally, we have shown that the SKA in its final (AA4) configuration could significantly improve the sensitivity over HERA to $\beta_\s$ and $N_{\rm eff}$, but not $\alpha_\s$, which is most accurately determined by CMB data.

Overall, our results demonstrate that 21\,cm surveys can provide a valuable complementary probe of features like departures from the minimal power-law description of the primordial power spectrum, and significantly reduce degeneracies found within CMB data. This again highlights the complementarity between these types of experiments.

Beyond constraints on the primordial power spectrum, we show (in App.~\ref{sec:appendix_growth_factor_transfer_function}) that using the Eisenstein \& Hu approximation for the transfer function of the linear power spectrum of matter perturbations and the Liddle approximation of the linear growth factor, which is the default in {\tt 21cmFAST}, significantly biases the output HMF when compared to using the more exact results from  {\tt CLASS}. This parasite effect can affect any analysis based on {\tt 21cmFAST}.

\begin{acknowledgments}
The authors would like to thank Dominic Agius, Bryan Zaldivar andSébastien Clesse for useful discussions and comments on the manuscript, as well as Claude AI for proofreading and coding support. LLH and JRS have been supported by the Fonds de la Recherche Scientifique F.R.S.- FNRS as a senior research associate and a doctoral research fellow respectively  and are members of BLU-ULB (Brussels Laboratory of the Universe, blu.ulb.be). They acknowledge the support of the FNRS research grant number J.0134.24, the ARC program of the Federation Wallonie-Bruxelles and the support of the IISN convention No. 4.4503.15. VP acknowledges the European Union's Horizon Europe research and innovation programme under the Marie Sk\l odowska-Curie Staff Exchange grant agreement no.\ 101086085 -- ASYMMETRY. This work received funding support from the European Research Council (ERC) under the European Union's HORIZON-ERC-2022 (grant agreement no.\ 101076865). Computational resources have been provided by the Consortium des Équipements de Calcul Intensif (CÉCI), funded by the Fonds de la Recherche Scientifique de Belgique (F.R.S.-FNRS) under Grant No. 2.5020.11 and by the Walloon Region of Belgium. 
\end{acknowledgments}

\appendix
\section{Characterisation of the Halo mass function}
\label{app:HMF}
\subsection{Multiplicity Functions}
\label{app:multiplfn}
The multiplicity functions of Angulo, Watson-$z$ and Yung all provide fits to the following functional form of the multiplicity function:
\begin{align}
    f(\sigma)=A\left[\left(\frac{\beta}{\sigma}\right)^\alpha+1\right]^\delta e^{-\gamma/\sigma^2}\,. \label{eq:f_general}
\end{align}
In the Angulo fit, $A=0.201$, $\alpha=1$, $\beta=2.08$, $\gamma=1.172$ and $\delta=1.7$. In the redshift-dependent Watson-$z$ fit, $\gamma=1.318$ and $\delta=1$ remain fixed and
\begin{align}
    A(z)&=\Omega_{\rm m}(z)\left[0.990\times(1+z)^{-3.216}+0.074\right] \nonumber \\
    \alpha(z)&=\Omega_{\rm m}(z)\left[5.907\times(1+z)^{-3.599}+2.344\right] \nonumber \\
    \beta(z)&=\Omega_{\rm m}(z)\left[3.136\times(1+z)^{-3.058}+2.349\right]\,,
\end{align}
where $\Omega_{\rm m}(z)$ is the matter content of the universe at a given redshift. Finally, the Yung parametrisation was fitted to N-body simulations performed between $z=6$ and $z=19$, resulting in
\begin{align}
    A & = 0.13765772 - 0.01003821\times z + 0.00102964\times z^2 \nonumber \\
    \alpha & = 1.06641384 + 0.02475576\times z - 0.00283342 \times z^2 \nonumber \\
    \beta & = 4.86693806 + 0.09212356\times z - 0.01426283 \times z^2 \nonumber \\
    \gamma & = 1.19837952 - 0.00142967\times z - 0.00033074 \times z^2
\end{align}
with $\delta=1$. Because the Yung fit becomes ill-defined at $z>20$, we freeze the parameters' redshift evolution for $z>19$. Because the sensitivity of HERA and SKA significantly deteriorates above $z\gtrsim16$ (see Fig.~\ref{fig:21cm_ps_alpha_s_impact}), we do not expect this to affect our results.

\subsection{Impact of Choice of Growth Factor and Transfer Function on the HMF}
\label{sec:appendix_growth_factor_transfer_function}

\begin{figure}[h]
    \centering
    \includegraphics[width=0.9\linewidth]{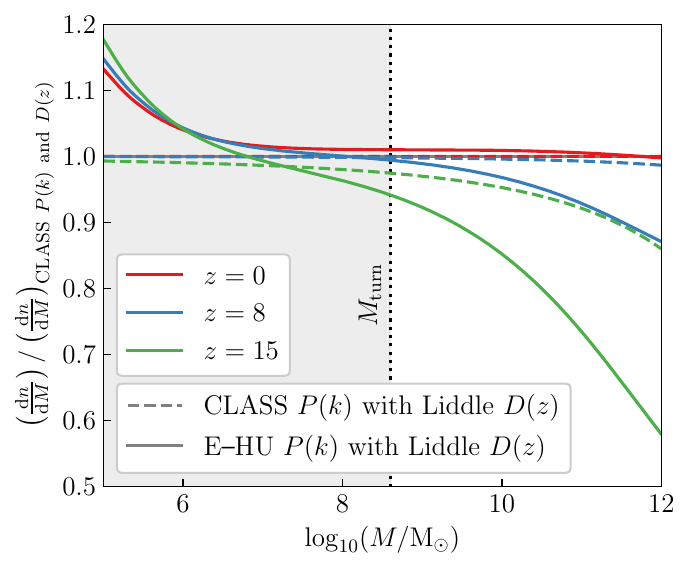}
    \caption{{\bf Dependence of the HMF on the  transfer function $T(k)$ and the growth factor $D(z)$.} \textit{Solid lines:} Ratio of the HMF using Eisenstein \& Hu approximation for $T(k)$ in conjunction with the Liddle  $D(z)$ approximation, as is the default in \texttt{21cmFAST}, to the HMF using $T(k)$ and $D(z)$ from {\tt CLASS}. \textit{Dashed lines:} Ratios of the Sheth--Tormen HMF computed using the Liddle $D(z)$ approximation only to that computed using the exact solution for $D(z)$ from \texttt{CLASS}. Shown are redshifts $z=0$ (red), $z=8$ (blue) and $z=15$ (green).  Even though the disagreement between the different methods is minor at $z=0$, the discrepancy becomes severe at higher redshifts.}
    \label{fig:ehu_class_hmf_comparison}
\end{figure}

By default, \texttt{21cmFAST} uses the Eisenstein \& Hu approximation~\cite{Eisenstein:1997jh} for the transfer function $T(k)$ and the Liddle~\cite{Liddle:1995pd} approximation for the growth factor $D(z)$ in the computation of Eq.~(\ref{eq:sigma_matter}). Given the exponential dependency of the multiplicity function of Eqs.~(\ref{eq:f_st}) and~(\ref{eq:f_general}) on the redshift-dependent matter variance, small errors in $T(k)$ or $D(z)$ may lead to significant biases in the amplitude of the HMF. 

This is illustrated with the continuous lines of Fig.~\ref{fig:ehu_class_hmf_comparison}, that compare the default (Sheth--Tormen) HMF treatment in {\tt 21cmFAST}, which uses both the Liddle $D(z)$ approximation and Eisenstein \& Hu approximation for $T(k)$, to the HMF as calculated using the transfer function and growth factor sourced from {\tt CLASS}. We show redshifts $z=0$ (red), $z=8$ (blue) and $z=15$ (green). Using both approximations, the amplitude of the HMF becomes artificially suppressed at $M\gtrsim10^8\,{\rm M}_\odot$ and artificially enhanced at $M\lesssim10^7\,{\rm M}_\odot$ compared to the {\tt CLASS} HMF. The low-mass enhancement is sourced by baryon pressure effects, which suppress the linear matter power spectrum at $k\gtrsim100\,{\rm Mpc}^{-1}$ and are not captured by the Eisenstein \& Hu approximation (for more details, see e.g.~\cite{Lewis:2007kz} and Appendix~A therein). 

The dashed lines show the impact of using the Liddle $D(z)$ approximation only (while the transfer function is  fed from {\tt CLASS}). While the HMFs in both cases are by definition identical at $z=0$, some discrepancy arises at higher redshifts. Yet, comparing the dashed and solid lines, we see that the larger source of bias in the HMF at large halo masses arises from the poor choice of transfer function.

To avoid biases in the imprints of both astrophysical and cosmological parameters, we therefore use \texttt{CLASS} to compute both the growth factor and input transfer function for each cosmology. The preceding discussion is however relevant for any $21\,$cm analyses based on {\tt 21cmFAST}, beyond just the impact of the running of the PPS.

\section{$\Lambda$CDM Forecast}

\begin{figure}[t]
    \centering
    \includegraphics[width=\linewidth]{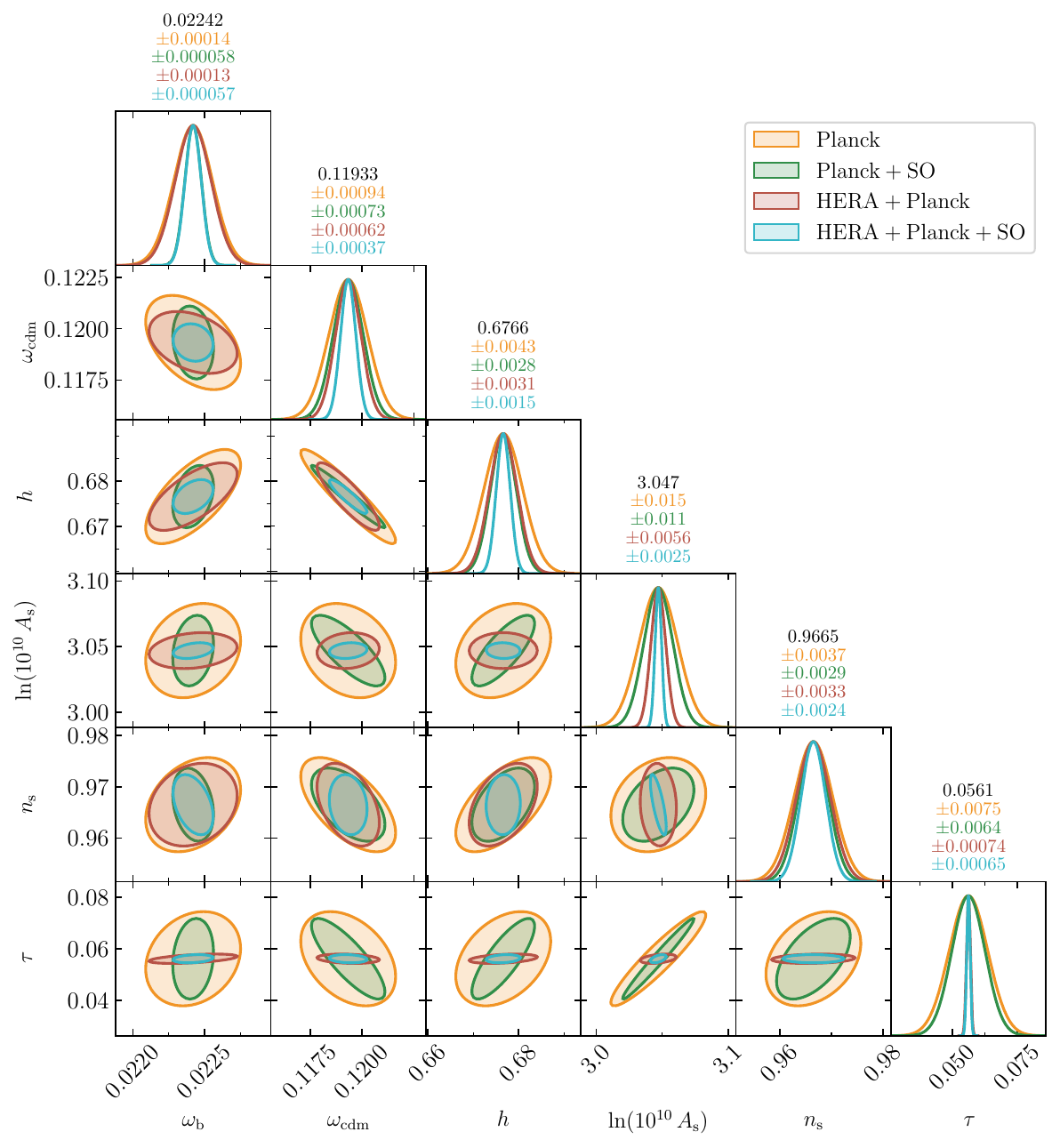}
    \caption{Same as Fig.~\ref{fig:alpha_s_so}, but showing the full set of $\Lambda$CDM parameters with $\alpha_\s=\beta_\s=0$ for Planck data alone (orange), the combination of Planck and SO data (green), the combination of HERA and Planck data (terracotta) and the full combination of HERA, Planck and SO data (blue).}
    \label{fig:lcdm_triangle_plot}
\end{figure}

Figure~\ref{fig:lcdm_triangle_plot} shows a triangle plot comparing Planck measurements (orange) to forecasts of the combinations of Planck and SO data (green), HERA and Planck data (terracotta) and HERA, Planck and SO data (blue) for the base $\Lambda$CDM model, i.e. $\alpha_\s=\beta_\s=\Delta N_{\rm eff}\equiv0$. This illustrates how synergies between Planck and SO CMB data as well as HERA $21\,$cm observations allow to set significantly tighter constraints on the parameters of the $\Lambda$CDM model. In particular, the much better determination of $\tau$ by HERA than from the CMB low-$\ell$ polarisation tail leads to significantly improved constraints on $A_\s$, $\omega_{\rm cdm}$ and $h$ as well. We particularly point out the excellent agreement between our HERA+Planck forecast and that of~\cite{Liu:2015txa}, Tab.~II (keeping in mind that~\cite{Liu:2015txa} used older Planck data).

\section{\label{sec:appendix}Triangle Plots}

\begin{figure*}
    \centering
    \includegraphics[width=0.99\linewidth]{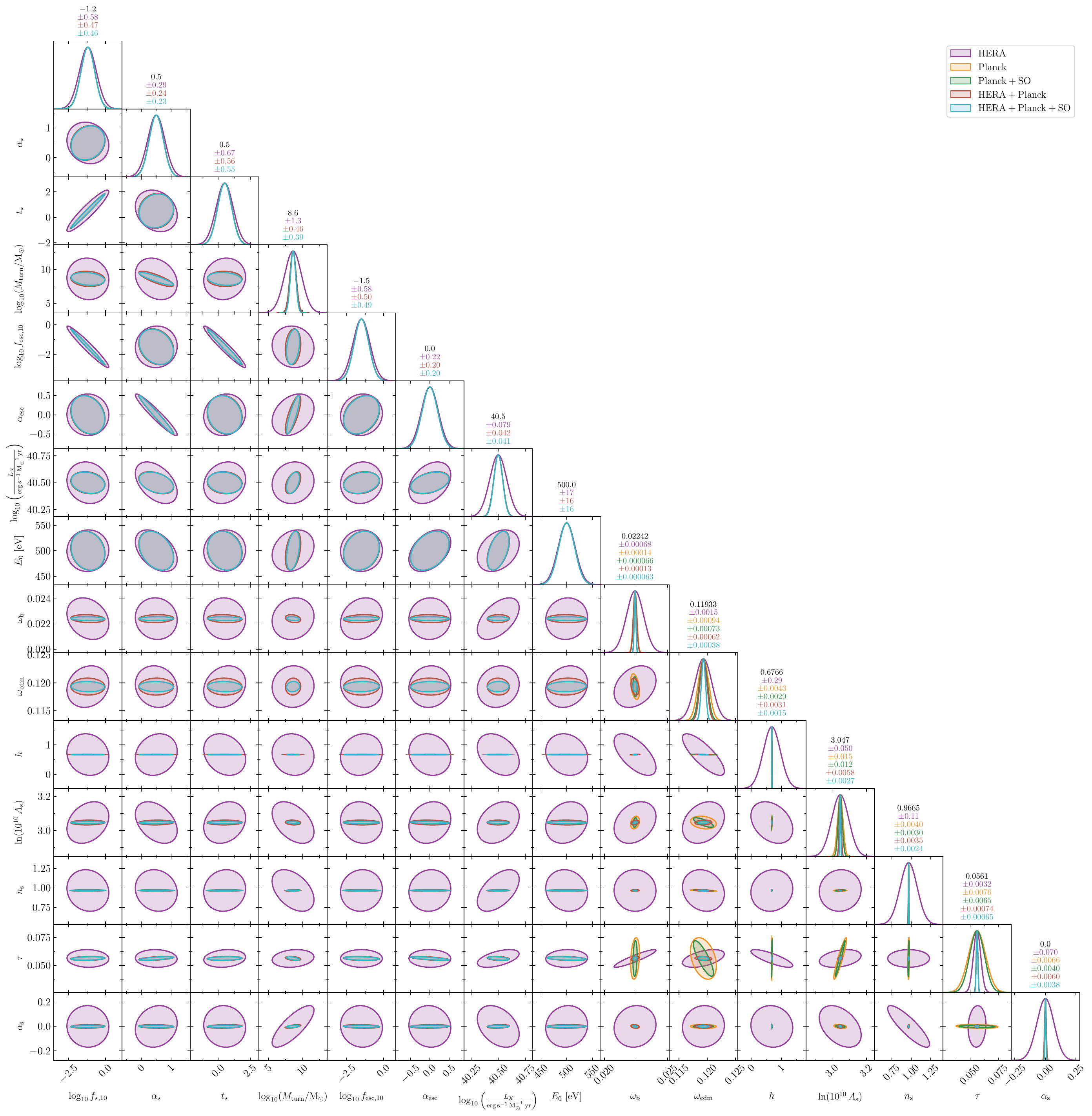}
    \caption{Same as Figs.~\ref{fig:alpha_s_hera_planck} and~\ref{fig:alpha_s_so}, but showing all parameters considered in the Fisher forecast.}
    \label{fig:alpha_s_large}
\end{figure*}

\begin{figure*}
    \centering
    \includegraphics[width=0.99\linewidth]{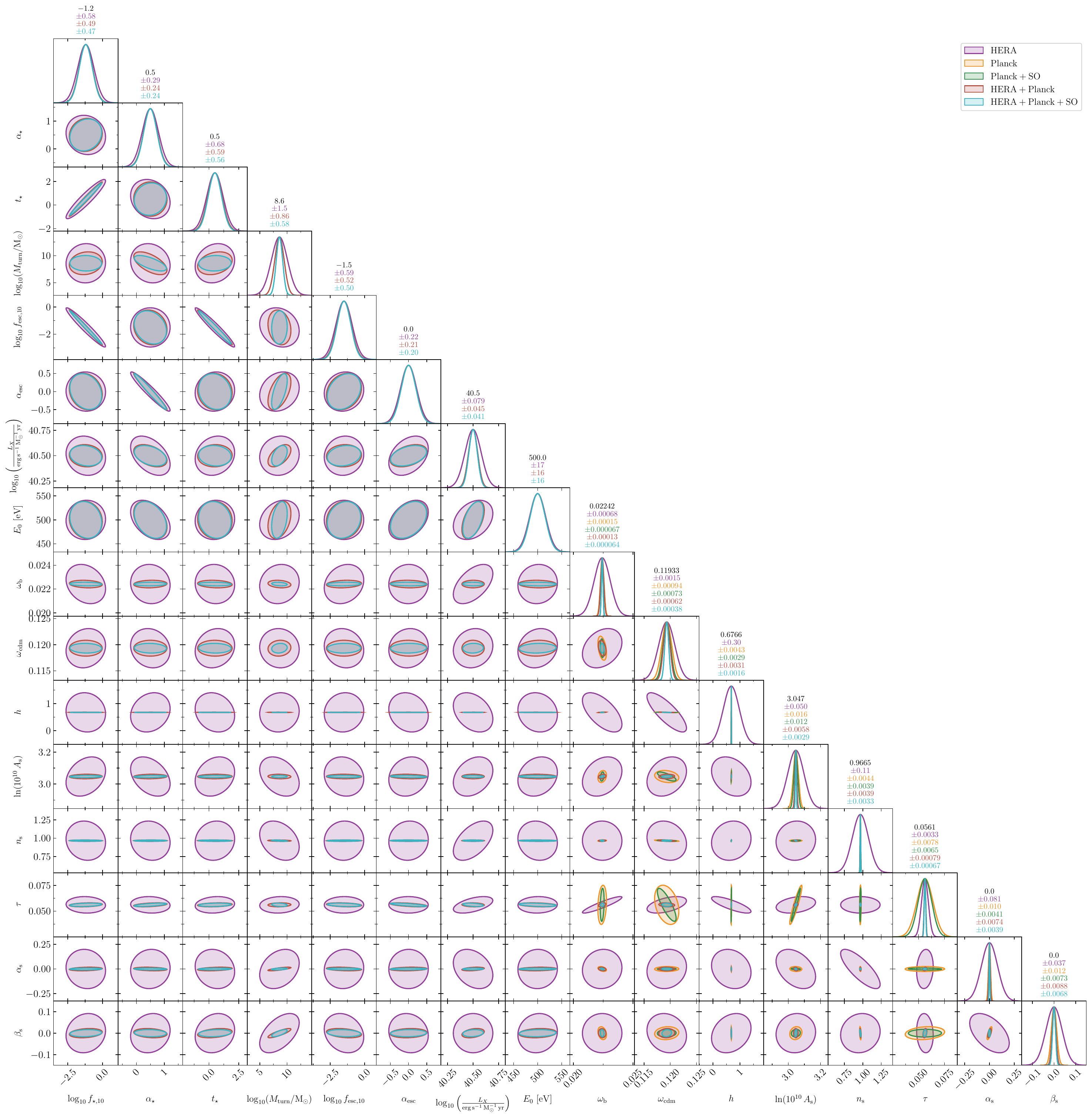}
    \caption{Same as Figs.~\ref{fig:beta_s_hera_planck} and~\ref{fig:beta_s_so}, but showing all parameters considered in the Fisher forecast.}
    \label{fig:beta_s_large}
\end{figure*}

\begin{figure*}
    \centering
    \includegraphics[width=0.99\linewidth]{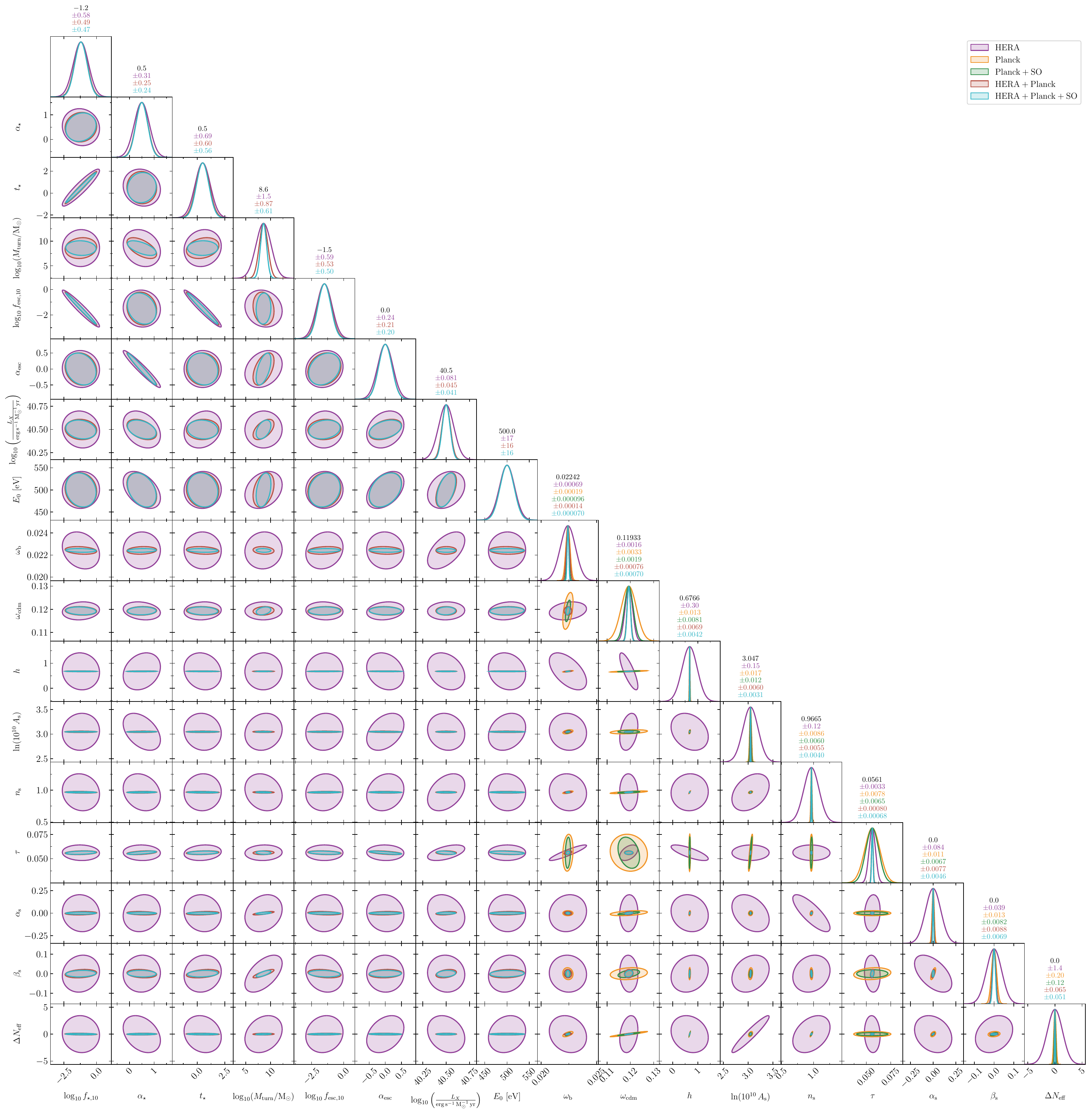}
    \caption{Same as Fig.~\ref{fig:neff}, but showing all parameters considered in the Fisher forecast, as well as the posterior from HERA alone.}
    \label{fig:neff_large}
\end{figure*}

Figure~\ref{fig:alpha_s_large} shows the marginalised posteriors of the $\bm\theta_{\alpha_\s}$ model from our HERA $21\,$cm forecast, Planck data, our Planck+SO forecast, as well as our HERA+Planck and HERA+Planck+SO forecasts, respectively. Figs.~\ref{fig:beta_s_large} and~\ref{fig:neff_large} show the posteriors of the $\bm\theta_{\beta_\s}$ and $\bm\theta_{\Delta N_{\rm eff}}$ models for the same dataset combinations respectively.

\bibliography{apssamp}

\end{document}